\documentclass[a4paper,11pt]{article}
\pdfoutput=1
\usepackage{jcappub}
\usepackage{tikz,xcolor,hyperref}
\usepackage{amsmath, amssymb, amsthm, graphicx, epsfig, fancyhdr,epsfig, slashed}
\usepackage[normalem]{ulem}
\usepackage{tikzsymbols}
\usepackage{natbib}
\usepackage{float}
\usepackage{placeins}
\usepackage[normalem]{ulem}

\newcommand{\gs}{g_\star}
\newcommand{\gss}{g_{\star s}}
\newcommand{\Trh}{T_\text{rh}}
\newcommand{\arh}{a_\text{rh}}
\newcommand{\Tmax}{T_\text{max}}
\newcommand{\rR}{\rho_R}
\newcommand{\rp}{\rho_\phi}

\newcommand{\DNeff}{\Delta N_\text{eff}}

\newcommand{\aend}{a_\text{end}}
\newcommand{\mend}{m_\text{end}}
\newcommand{\rend}{\rho_\text{end}}

\title{Is leptogenesis during gravitational reheating flavourful?}
\author[a]{Basabendu Barman,}
\author[b]{Arghyajit Datta,}
\author[c]{and Md Riajul Haque}
\affiliation[a]{\,\,Department of Physics, School of Engineering and Sciences,
SRM University AP, Amaravati 522240, India}
\affiliation[b]{\,\,Laboratory for Symmetry and Structure of the Universe, Department of Physics,
Jeonbuk National University, Jeonju 54896, Republic of Korea}
\affiliation[c]{\,\,Physics and Applied Mathematics Unit, Indian Statistical Institute, 203 B.T. Road, Kolkata 700108, India}
\emailAdd{basabendu.b@srmap.edu.in}
\emailAdd{arghyad053@gmail.com}
\emailAdd{riaj.0009@gmail.com}
\abstract{We examine the impact of charged lepton Yukawa equilibration on leptogenesis during gravitational reheating. During the post-inflationary era, the inflaton field is assumed to oscillate around the minimum of a monomial potential, leading to the gravitational production of Standard Model (SM) particles, constituting the radiation bath. The heavy right-handed neutrinos (RHN), responsible for generating baryon asymmetry via leptogenesis, are also produced through graviton-mediated scattering of the homogeneous inflaton field and thermal bath, as well as from the the inverse decay of the bath particles. By considering both minimal and non-minimal gravitational contributions to SM, we demonstrate that flavour effects can be safely neglected in the minimal reheating scenario. However, with large non-minimal coupling, these effects may become important, depending on the choice of the RHN mass. We identify the corresponding viable parameter space that satisfies the observed baryon asymmetry in each case.
}
\begin{document}
\maketitle
\section{Introduction}
\label{sec:intro}
The observed matter-antimatter asymmetry of the Universe is a longstanding puzzle in particle cosmology. While the hot Big Bang model successfully predicts the cosmic microwave background (CMB) radiation and the primordial abundances of light elements, it fails to explain why the cosmic baryon-to-photon ratio is $\eta\equiv n_B/n_\gamma\sim 6\times 10^{-10}$~\cite{Planck:2018vyg}. Given that the Universe is expected to have originated with equal amounts of matter and antimatter, and any initial asymmetry would have been diluted by the accelerated expansion during inflation, a dynamical mechanism is required to account for the observed asymmetry. This mechanism must satisfy the Sakharov conditions~\cite{Sakharov:1967dj}, which demand (i) baryon number $B$ violation, (ii) violation of $C$ and $CP$ symmetries, and (iii) a departure from thermal equilibrium. While the Standard Model (SM) of particle physics includes the first two ingredients, the out of equilibrium condition can not be satisfied due to the  SM Higgs mass (see, for example, Refs.~\cite{Carena:1997ys,Trodden:1998qg,Cline:2006ts,Morrissey:2012db} for detailed discussions). Consequently, the observed asymmetry can not be addressed within the framework of the SM. A well-known approach to baryogenesis involves the out-of-equilibrium decay of a beyond SM heavy particle that generates the baryon asymmetry of the Universe (BAU)~\cite{PhysRevLett.42.850,Kolb:1979qa}\footnote{Ref.\cite{Flores:2024lzv} investigate the generation of the baryon asymmetry through gravitational particle production, employing the Bogoliubov approach.}. One particularly interesting implementation of this mechanism is leptogenesis~\cite{Fukugita:1986hr}, in which a net leptonic asymmetry is first generated, which is then converted into baryon asymmetry via $B+L$-violating electroweak sphaleron transitions~\cite{Kuzmin:1985mm}. Notably, in this case, the necessary lepton asymmetry can arise from CP-violating out-of-equilibrium decays of the same heavy particles involved. In the case of a conventional Type-I seesaw scenario~\cite{Minkowski:1977sc,Yanagida:1979as,Yanagida:1979gs,GellMann:1980vs,Mohapatra:1979ia,Schechter:1980gr,Schechter:1981cv,Datta:2021elq}, the heavy particles\footnote{Ref.~\cite{Domcke:2020quw} discusses ``wash-in" leptogenesis, where the RHNs can have masses as low as a few 100 TeV.} could be the right-handed neutrinos (RHN)~\cite{Luty:1992un,Pilaftsis:1997jf,Barbieri:1999ma}. Such a setup also provides an explanation for the smallness of neutrino masses—another phenomenon that the SM fails to explain. 

The dynamical generation of baryon asymmetry through leptogenesis has predominantly been studied within a radiation-dominated background (see, for instance, Refs.~\cite{Davidson:2008bu, Buchmuller:2004nz} for a review). However, as it is well known, an elegant solution to the cosmological puzzles, namely, the horizon and flatness problems~\cite{Guth:1980zm, Linde:1981mu} arises by positing an inflationary phase in the very early Universe. In its simplest avatar, inflation can be modeled by a single, slowly rolling scalar field with an almost flat potential, which dominates the energy density of the primordial Universe. This phase of cosmic inflation dilutes any pre-existing matter and radiation, necessitating a {\it reheating} mechanism to transition the Universe to a radiation-dominated state. Consequently, it is highly plausible that the observed baryon asymmetry arose dynamically from the CP-violating decay of right-handed neutrinos produced during this reheating epoch. Such productions are interesting as they crucially depend on the reheating dynamics themselves. Depending on the hierarchy between the RHN mass, the reheating temperature $\Trh$ and the maximum temperature of the thermal bath $\Tmax$ during reheating, one can either have thermal (typically when $\Trh$ is above the RHN mass) or non-thermal leptogenesis~\cite{Lazarides:1991wu,Kumekawa:1994gx,Asaka:1999yd,Lazarides:1999dm,Giudice:1999fb,Asaka:1999jb,Hamaguchi:2001gw,Buchmuller:2004tu,Senoguz:2007hu,Hahn-Woernle:2008tsk,Baer:2008eq,Fukuyama:2010hh,Datta:2020bht,Barman:2021tgt,Asaka:2002zu,Ghoshal:2022fud,Datta:2022jic,Samanta:2020gdw,Datta:2023pav,Zhang:2023oyo,Granelli:2023egb,Bhandari:2023wit,Datta:2024tne}.

With this motivation, in this work, we have discussed the generation of baryon asymmetry via leptogenesis during the period of reheating. As a concrete model, we have considered that the inflaton oscillates at the bottom of a monomial potential during reheating,  where the radiation bath is produced from the scattering of inflaton condensate, mediated by massless gravitons, leading to what is known as gravitational reheating~\cite{Clery:2021bwz,Haque:2022kez}. Since gravitational coupling is universal and unavoidable, we consider such gravity-mediated production to apply to the RHNs as well, where they are produced gravitationally from scattering of the inflaton condensate, as well as the particles in the thermal bath. Such a scenario, dubbed as gravitational leptogenesis, has been studied, for example, in~\cite{Co:2022bgh,Barman:2022qgt,Clery:2023mjo,Haque:2023zhb,Barman:2023opy,Barman:2024slw}. Here we are rather interested in the equilibration of individual charged lepton Yukawa interactions during gravitational reheating scenario, that can potentially lead to {\it flavour effects}~\cite{Nardi:2005hs,DeSimone:2006nrs,Nardi:2006fx,Blanchet:2006be,Blanchet:2006ch,Davidson:2008bu,Blanchet:2011xq,Moffat:2018wke,Granelli:2021fyc} during the lepton asymmetry generation. Within the framework of gravitational reheating, the asymmetry calculations have been typically performed under one-flavour approximation, assuming that the final total $B-L$ asymmetry is not sensitive to the dynamics of the individual $B/3-L_\alpha$ asymmetries. Nevertheless, the lepton doublets of different flavours are distinguished by their Yukawa couplings $y_\alpha$ associated to the interaction $y_\alpha \bar{\ell}_L H E_R$ in the Lagrangian\footnote{Here we only consider flavour effects due to the SM charged leptons; however, as discussed in~\cite{Dev:2017trv}, coherences amongst the heavy-neutrino flavours have an important effect on the source of
CP asymmetry due to oscillations.}. During leptogenesis, these differences become relevant if the interactions mediated by $y_\alpha$ are faster than the leptogenesis process and the expansion rate of the Universe. Since the Yukawa couplings for different flavours $e, \mu, \tau$ differ significantly in strength, they induce variations in the interaction rates associated with the charged lepton Yukawa coupling. Addressing this effect requires recognizing that when (some of) these charged lepton Yukawa interactions are ``fast enough," and accordingly constructing the Boltzmann equation(s) (BEQ) for the relevant lepton flavour(s). Conversely, when the charged lepton Yukawa interactions are much slower than the expansion rate of the Universe, leptogenesis remains insensitive to lepton flavours, necessitating a modification of the BEQ to account for the unflavoured or one-flavour approximation.

The paper is organized as follows. In Sec.~\ref{sec:reheat}, we discuss the gravitational reheating scenario in detail. Then, we move on to the discussion of leptogenesis during reheating in Sec.~\ref{sec:lepto}, elaborating on our findings. Finally, we conclude in Sec.~\ref{sec:concl}.
\section{The background dynamics}
\label{sec:reheat}
Following the single-field inflationary paradigm, the energy density of the Universe at the end of inflation is dominated by the energy density of a real scalar field $\phi$. For the scalar field potential $V(\phi)$, we consider the class of $\alpha$-attractor T-models \footnote{Note that the assumption of the power law behavior of the potential from the starting point of reheating is only valid under the approximation $\phi_{\rm end}/M_{\rm P}<1$. However, with increasing $k$ value, such approximation is not well justified at the starting point of reheating as  $\phi_{\rm end}$ becomes (sub-) Planckian. To avoid such an issue, we restrict ourselves to $k\leq 20$.}~\cite{Kallosh:2013hoa}
\begin{align}
V(\phi)=\lambda M_P^{4}\left[\sqrt{6} \tanh \left(\frac{\phi}{\sqrt{6} M_P}\right)\right]^{k}=
\lambda M_P^{4}\times
\begin{cases}
1 & \; \text{for}\; \phi \gg M_P,
\\
\left(\frac{\phi}{M_P}\right)^k & \; \text{for}\; \phi\ll M_P\,.
\end{cases}
\end{align}
The overall scale of the potential parameterized by the coupling $\lambda\simeq 18\pi^2 A_S/\left(6^{k/2}\,N_\star^2\right)$ is determined from the amplitude of the CMB power spectrum $\ln(10^{10}A_S)=3.044$~\cite{Planck:2018jri} and the number of e-folds $N_\star$ measured from the end of inflation to the time when the pivot scale $k_*=0.05~{\rm Mpc}^{-1}$ exits the horizon. The inflaton mass, in this case, takes the form
\begin{align}\label{eq:inf-mass}
& m_\phi(a)=\sqrt{\lambda\, k\,(k-1)}\,\left(\frac{\rho_\phi}{\lambda\,M_P^4}\right)^\frac{k-2}{2k}
\implies m_\phi(a)=m_{\rm end}\,\left(\frac{\rp}{\rend}\right)^\frac{k-2}{2k}\,,
\end{align}
where $m_{\rm end}=m_\phi(\aend)$ is the inflaton mass at the end of inflation. Note that, as expected, for $k=2$, the inflaton mass is constant. In order to track the evolution of the inflaton $(\rp)$ and radiation $(\rR)$ energy densities, we solve the following set of coupled BEQ~\cite{Giudice:2000ex}
\begin{align}
    &\frac{d\rho_\phi}{dt} + 3H\,(1+w_\phi)\, \rho_\phi = -(1+w_\phi)\,\Gamma_\phi^h\, \rho_\phi\,,\label{eq:BErhop}\\
    &\frac{d\rho_R}{dt} + 4H\, \rho_R = + (1+w_\phi)\,\Gamma_\phi^h\, \rho_\phi\,, 
    \label{eq:BErhoR}
\end{align}
together with 
\begin{align}\label{eq:hub1}
H = \sqrt{\frac{\rho_\phi+\rR}{3\,M_P^2}}\,.    
\end{align}
We define $w_\phi\equiv p_\phi/\rho_\phi=(k-2)/(k+2)$~\cite{Garcia:2020wiy} as the general equation of state (EoS) parameter for the background\footnote{As discussed in~\cite{Lozanov:2016hid,Lozanov:2017hjm,Maity:2018qhi}, for any potential steeper than quadratic near the origin, the oscillating inflaton condensate can fragment due to self-resonance. In this case, the equation of state approaches $w \to 1/3$ at sufficiently late times. If this happens after $T = \Trh$, the effect is negligible because the inflaton energy would already be subdominant. Moreover, since RHNs are primarily produced at $\aend$, any fragmentation occurring later would not affect the baryon asymmetry calculation. While reheating could be influenced, the exact moment when $w$ transitions to $1/3$ depends on $k$ and would require dedicated lattice simulations. Self-resonance becomes less efficient for larger $k$, as shown by lattice results for $k$ up to 6~\cite{Lozanov:2016hid,Lozanov:2017hjm,Maity:2018qhi}, whereas most of our viable results apply to $k > 6$. Conducting such lattice simulations for higher $k$ is beyond the scope of this study.}. As during major part of the reheating, the total energy density of the Universe is dominated by the inflaton, the expansion rate corresponding to the term $3\,H\,(1+w_\phi)\,\rho_\phi$ dominates over the reaction rate $(1+w_\phi)\,\Gamma_\phi\rho_\phi$ in Eq.~\eqref{eq:BErhop}. It is then possible to solve Eq.~\eqref{eq:BErhop} analytically, obtaining
\begin{equation}
\rho_\phi(a) \simeq \rho_{\rm end}\,\left(\frac{a_{\rm end}}{a} \right)^\frac{6\,k}{k+2}\,,
\label{eq:rhoPh}
\end{equation}
with corresponding Hubble rate
\begin{align}\label{eq:hubble}
& H(a)\simeq H_{\rm end}\,\left(\frac{a_{\rm end}}{a} \right)^\frac{3\,k}{k+2}\,.  
\end{align}  
Here $\rend$ is the energy density of the Universe at the end of inflation, that reads
\begin{align}\label{eq:rho-end}
& \rend=\frac{3}{2}\,V(\phi_{\rm end})\approx\frac{3}{2}\,\lambda\,M_P^4\,\left(\phi_{\rm end}/M_P\right)^k\,,    
\end{align}
with
\begin{align}
& H_{\rm end}\simeq M_P\,\sqrt{\lambda/2}\,\left(\phi_{\rm end}/M_P\right)^{k/2}\,,   
\end{align}
where $\phi_{\rm end}$ is the value of the inflaton field at the end of inflation. This can be obtained from the condition $\epsilon_V\simeq 1$, leading to
\begin{align}\label{eq:phi-end}
&\phi_{\rm end}\simeq \sqrt{\frac{3}{8}}\,M_P\,\log \left(\frac{1}{2}+\frac{k}{3}\,\left[k+\sqrt{k^2+3}\right]\right)\,,    
\end{align}
where $\epsilon_V=\left(M_P^2/2\right)\,\left(V'(\phi)/V(\phi)\right)^2$\, is the potential slow-roll parameter~\cite{Baumann:2022mni}. Using the expression in Eq.~\eqref{eq:rho-end}, we find, for $k=10$, $\rend\simeq 10^{64}\,\text{GeV}^4$.   
\begin{figure}[htb!]
\centering
\includegraphics[scale=0.13]{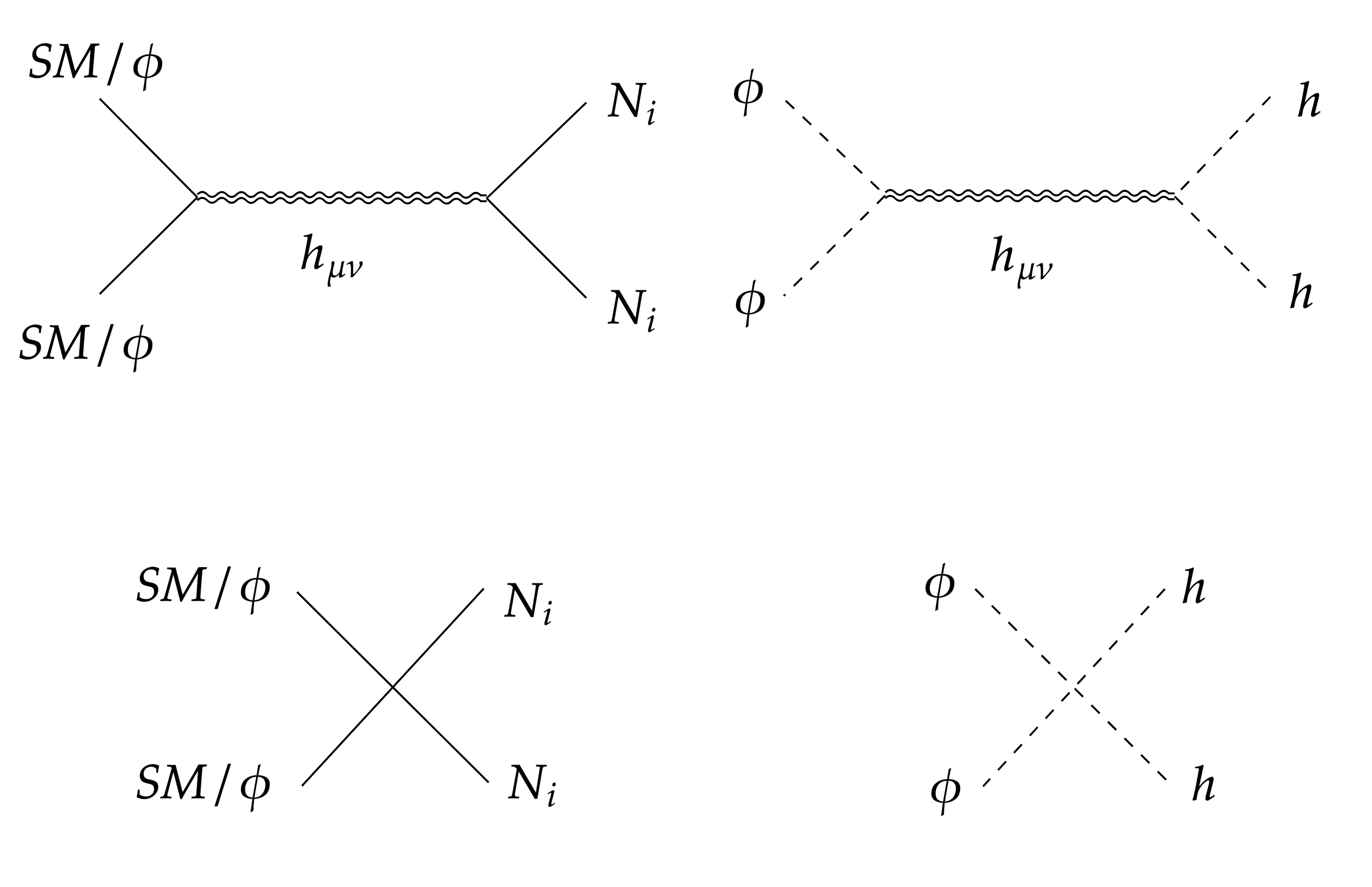}
\caption{Minimal (top) and non-minimal (bottom) gravitational production of the RHNs $N_i$ and SM particles.}
\label{fig:feyn}
\end{figure}

We are interested in purely gravitational production of both the SM and beyond the SM fields, where the interactions are mediated by gravity. This can be realized by introducing perturbations around the flat metric: $g_{\mu \nu}\simeq \eta_{\mu \nu}+2h_{\mu \nu}/M_P;\,|h_{\mu\nu}|\ll\eta_{\mu\nu}$, where $h_{\mu\nu}$ is the rank-2 tensorial graviton field. The gravitational interactions are then given by~\cite{Choi:1994ax,Holstein:2006bh}
\begin{align}
\sqrt{-g}\,\mathcal{L}_{\rm int}= -\frac{1}{M_P}\,h_{\mu \nu}\,\left(T^{\mu \nu}_{\rm SM}+T^{\mu \nu}_\phi + T^{\mu \nu}_{X} \right) \,,
\label{Eq:lagrangian}
\end{align}
where $\phi$ is the inflaton, and $X$ is a particle that does not belong to the SM. In the present context, we consider $X$ to be a spin 1/2 Majorana fermion. The graviton propagator is defined as
\begin{equation}
\Pi^{\mu\nu\rho\sigma}(p) = \frac{1}{2p^2}\,\left(\eta^{\rho\nu}\eta^{\sigma\mu} + \eta^{\rho\mu}\eta^{\sigma\nu} - \eta^{\rho\sigma}\eta^{\mu\nu} \right) \, .
\end{equation}
The form of the stress-energy tensor $T^{\mu \nu}_i$ depends on the spin of the field and, for a Majorana fermion $\chi$,
\begin{equation}\label{eq:tmunu}
T^{\mu \nu}_{1/2} =
\frac{i}{8}
\left[ \bar \chi \gamma^\mu \overset{\leftrightarrow}{\partial^\nu} \chi
+\bar \chi \gamma^\nu \overset{\leftrightarrow}{\partial^\mu} \chi \right] -g^{\mu \nu}\left[\frac{i}{4}
\bar \chi \gamma^\alpha \overset{\leftrightarrow}{\partial_\alpha} \chi
-\frac{m_\chi}{2}\,\overline{\chi^c} \chi\right]\,,
\end{equation}
whereas for a scalar $S$,
\begin{equation}
T^{\mu \nu}_{0} =
\partial^\mu S \partial^\nu S-
g^{\mu \nu}
\left[
\frac{1}{2}\partial^\alpha S\,\partial_\alpha S-V(S)\right]\,.
\label{eq:tmunuphi}
\end{equation}
The production rate of radiation, in this case, is given by~\cite{Clery:2021bwz, Clery:2022wib, Co:2022bgh}
\begin{equation}
   (1+w_\phi)\,\Gamma_\phi^h\, \rho_\phi = R^{\phi^k}_H \simeq \frac{\rho_{\phi}^2}{4\,\pi M_P^4} \sum_{n=1}^{\infty}  2n\omega|{\mathcal{P}}^k_{2n}|^2 = \alpha_k M_P^5 \left(\frac{\rho_{\phi}}{M_P^4}\right)^{\frac{5k-2}{2k}}\,.   
   \label{Eq:ratephik}
\end{equation}
In evaluating the interaction rate, the inflaton is treated as an oscillating classical field, which is parametrized as
\begin{equation}
    \label{Eq:oscillation}
    \phi(t)= \phi_0(t)\times\mathcal{T}(t) = \phi_0(t)\sum_{n=-\infty}^{\infty}\,{\cal T}_n\,e^{-in \omega t}\,,
\end{equation}
where $\phi_0(t)$ is the time-dependent amplitude that includes the effects of redshift and $\mathcal{T}(t)$ describes the periodicity of the oscillation. We also expand the potential energy in terms of the Fourier modes as~\cite{Ichikawa:2008ne,Kainulainen:2016vzv,Clery:2021bwz,Co:2022bgh,Garcia:2020wiy,Ahmed:2022qeh}
\begin{align}
V(\phi)=V(\phi_0)\sum_{n=-\infty}^{\infty} {\cal P}_n^ke^{-in \omega t}
=\rho_\phi\sum_{n = -\infty}^{\infty} {\cal P}_n^ke^{-in \omega t}\,,
\end{align}
where the frequency of oscillations of $\phi$ field reads~\cite{Garcia:2020wiy}
\begin{equation}
\label{eq:angfrequency}
\omega=m_\phi \sqrt{\frac{\pi k}{2(k-1)}}
\frac{\Gamma(\frac{1}{2}+\frac{1}{k})}{\Gamma(\frac{1}{k})}\,.
\end{equation}
By solving Eq.~\eqref{eq:BErhoR}, using the Eq.~\eqref{eq:rhoPh} together with the interaction rate~\eqref{Eq:ratephik}, one obtains the radiation energy density as
\begin{align}
    \rho_R(a) \simeq\sqrt{3}\,\alpha_k\,M_P^4\, \left(\frac{k+2}{8k-14}\right) \left(\frac{\rho_{\rm end}}{M_P^4}\right)^{\frac{2k-1}{k}} \left(\frac{a_{\rm end}}{a}\right)^4\left[1-\left(\frac{a_{\rm end}}{a}\right)^{\frac{8k-14}{k+2}}\right]\,.
    \label{eq:rhoR}
\end{align}
From Eq.~\eqref{eq:BErhop} and \eqref{eq:BErhoR}, we note that the thermalization process of the SM particles produced from the inflaton scattering helps the Universe to attain a maximum temperature $\Tmax$ right at the end of inflation. Subsequently, the temperature falls to $\Trh$, where equality between $\rho_\phi$ and $\rho_R$ is achieved. As a result, reheating temperature can be evaluated as
\begin{align}\label{eq:grav-trh}
& \Trh^4 = \frac{30}{\pi^2\,g_{\rm RH}}\, M_P^4\,\left(\frac{\rend}{M_P^4}\right)^{\frac{4k-7}{k-4}}\,\left(\frac{\alpha_k\,\sqrt{3}\,(k+2)}{8k-14}\right)^\frac{3k}{k-4}\,,
\end{align}
where $g_{\rm RH}$ is the relativistic degrees of freedom associated with the thermal bath calculated at the end of reheating.
\begin{table}[htb!]\large
\centering
\begin{tabular}{|c|c|c|}
\hline
$k$ & $\alpha_k$ & $\alpha_k^{\xi}$ \\
\hline
6 &  0.000193 & $\alpha_k + 0.00766 \, \xi^2$ \\
\hline
8 & 0.000528 & $\alpha_k + 0.0205 \, \xi^2$ \\
\hline
10 & 0.000966 & $\alpha_k + 0.0367\, \xi^2$ \\
\hline
12 & 0.00144 & $\alpha_k +  0.0537 \, \xi^2$ \\
\hline
14 &  0.00192 & $\alpha_k +  0.0702 \, \xi^2$ \\
\hline
16 &  0.00239 & $\alpha_k +  0.0855 \, \xi^2$ \\
\hline
18 &  0.00282 & $\alpha_k +  0.0995 \, \xi^2$ \\
\hline
20 &  0.00322 & $\alpha_k +  0.112 \, \xi^2$ \\
\hline
\end{tabular}
\caption{Relevant coefficients for the gravitational reheating [see.~Eq.~\eqref{Eq:ratephik} and Eq.~\eqref{ratexi}].}
\label{tab:alpha-k}
\end{table}
One can further note that, for $\aend\ll a\ll\arh$,  the temperature evolves as (see Eq.~\eqref{eq:rhoR}])
\begin{align}
& T(a)=\Trh\,\left(\frac{\arh}{a}\right)\,.
\end{align}
For $k > 4$, reheating from purely gravitational scattering is possible. However, it is very inefficient. One actually needs to go beyond $k=9$, for which the gravitational reheating temperature can be found to be $\Trh \simeq 2$ MeV. Before proceeding further, it is important to briefly address the need for non-minimal coupling. During inflation, quantum fluctuations naturally produce a scale-invariant spectrum of tensor metric perturbations on super-hubble scales. In the standard post-inflationary scenario, these tensor modes re-enter the horizon during the radiation-dominated (RD) era, and as a result, the gravitational waves spectrum turns out as scale invariant. However, it is well known that non-standard cosmologies—such as an early stiff epoch preceding radiation domination—disrupt this scale invariance in the primary gravitational wave spectrum. In such cases, the gravitational wave (GW) spectrum acquires a blue tilt at frequencies corresponding to modes that re-entered the horizon during the stiff era ($w_\phi>1/3$)~\cite{Giovannini:1998bp,Giovannini:1999bh,Riazuelo:2000fc,Seto:2003kc,Boyle:2007zx,Stewart:2007fu,Li:2021htg,Artymowski:2017pua,Caprini:2018mtu,Bettoni:2018pbl,Figueroa:2019paj,Opferkuch:2019zbd,Bernal:2020ywq,Ghoshal:2022ruy,Caldwell:2022qsj,Gouttenoire:2021jhk,Haque:2021dha,Chakraborty:2023ocr,Barman:2023ktz,Barman:2024slw,Maity:2024cpq}. As discussed in~\cite{Figueroa:2019paj,Ghoshal:2023phi,Barman:2023ktz,Chakraborty:2023ocr}\footnote{See ref.\cite{Bettoni:2024ixe} for Hubble-induced phase transition scenario as a source of a post-inflationary stochastic gravitational wave background.}, although a stiff phase in the expansion history significantly enhances the inflationary GWs background, potentially making it observable, it can also result in an excessive production of GW energy density violating the constraint from $\DNeff$ that encodes maximum allowed relativistic degrees of freedom around BBN and CMB. In our minimal gravitational reheating setup, we expect $w_\phi>0.65$ to get a reheating temperature above BBN energy scale, which is inconsistent with the $\DNeff$ bound, giving a GWs overproduction \cite{Haque:2022kez}. To make such a scenario consistent, one can increase the radiation energy density relative to the GW energy density by introducing a non-minimal contribution to the radiation, as demonstrated in Refs.~\cite{Figueroa:2018twl,Opferkuch:2019zbd,Barman:2022qgt,Barman:2023icn}. This provides a key motivation for introducing a non-minimally coupled scalar sector.

\begin{figure}[htb!]
    \centering
    \includegraphics[scale=0.37]{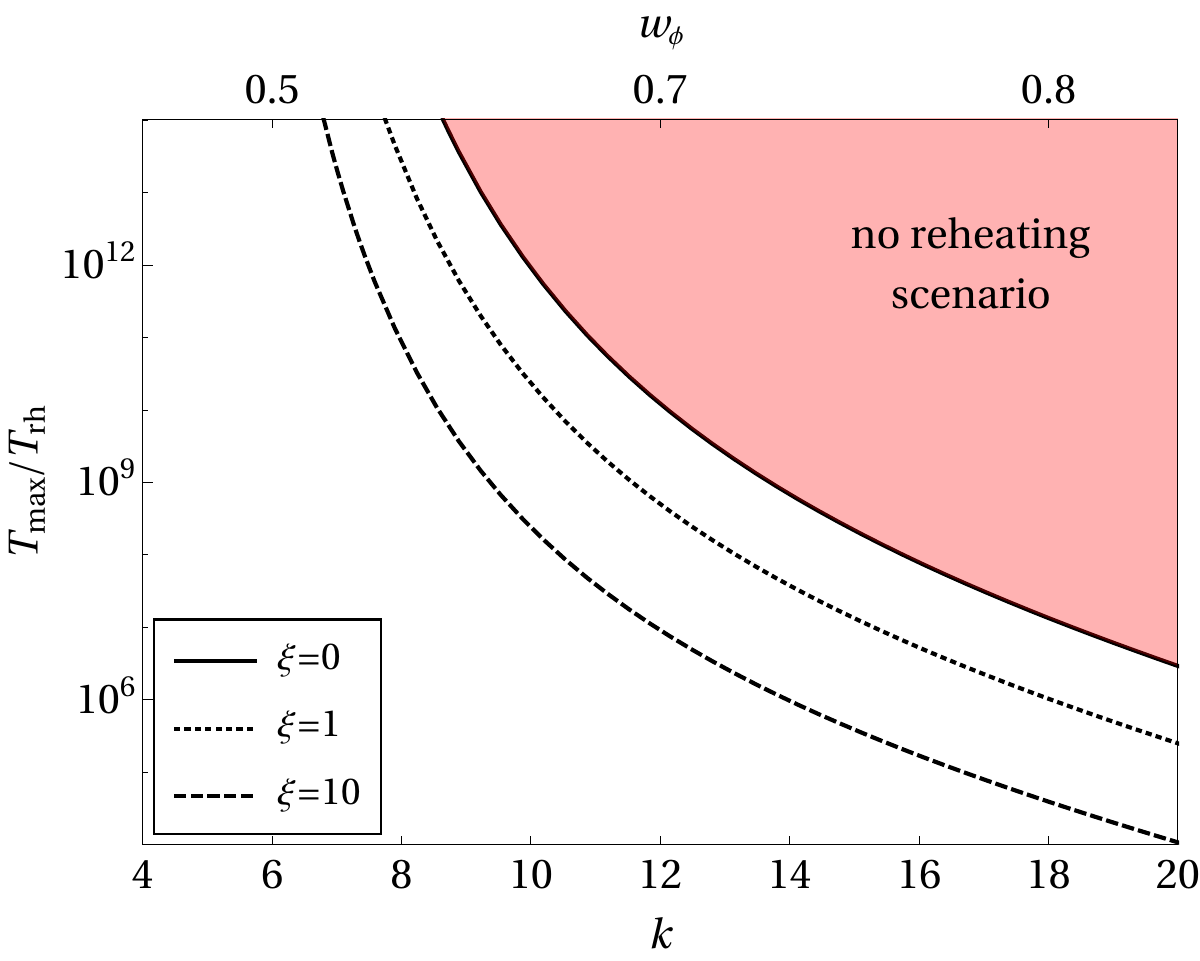}
    \includegraphics[scale=0.37]{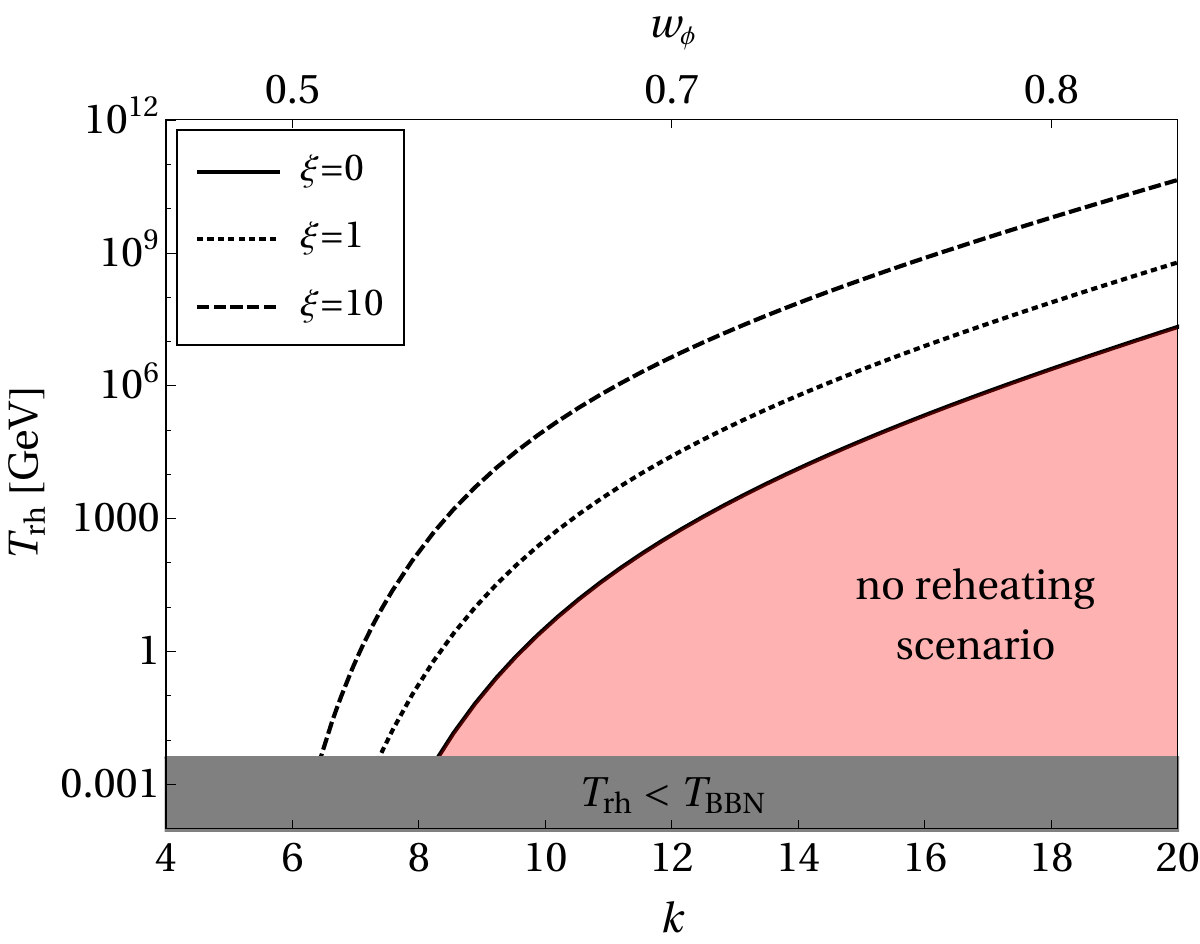}
    \caption{{\it Left:} Ratio of $\Tmax$ to $\Trh$ as a function of $k$ for minimal and non-minimal coupling scenarios. {\it Right:} $\Trh$ as a function of $k$, for the same set of $\xi$'s as in the left panel. The gray-shaded region is forbidden from the lower bound on $\Trh$ from BBN. Within the red shaded region, gravitational reheating is not possible (see text).}
    \label{fig:TRH}
\end{figure}

The non-minimal coupling of Higgs bosons to gravity provides an additional channel to reheat, with corresponding rate~\cite{Clery:2022wib,Co:2022bgh}
\begin{equation}
   (1+w_\phi)\,\Gamma_\phi \rho_\phi= R^{\phi,\xi}_H +R^{\phi^k}_H \simeq\alpha_k^{\xi}\,M_P^5\,\left(\frac{\rho_{\phi}}{M_P^4}\right)^{\frac{5k-2}{2k}}\,,
   \label{ratexi}
\end{equation}
where
\begin{equation}
R^{\phi,\xi}_H
\simeq \frac{\xi^2N_h}{4\pi M_P^4} \sum_{n=1}^{\infty}  2n\omega \left|2\times{\mathcal{P}}^k_{2n}\rho_{\phi} + \frac{(n\omega)^2}{2}\phi_0^2\,|\mathcal{T}_n|^2 \right|^2 \,.
\end{equation}
The numerical estimates of the co-efficient $\alpha_k^{\xi}$ for different values of $k$ are reported in Tab.~\ref{tab:alpha-k}. In the  non-minimal gravitational reheating setup, reheating temperature  can be obtained as 
\begin{align}\label{eq:non-minimal-trh}
& \left(\Trh^{\xi}\right)^4 = \frac{30}{\pi^2\,g_{\rm RH}}\, M_P^4\,\left(\frac{\rend}{M_P^4}\right)^{\frac{4k-7}{k-4}}\,\left(\frac{\alpha_k^{\xi}\,\sqrt{3}\,(k+2)}{8k-14}\right)^\frac{3k}{k-4}\,,
\end{align}
while the maximum temperature in this case reads,
\begin{align}\label{eq:tmax-xi}
& \left(\Tmax^{\xi}\right)^4=\frac{30\,\sqrt{3}}{\pi^2\,\gs}\,\alpha_k^{\xi}\,M_P^4\, \left(\frac{\rend}{M_P^4}\right)^\frac{2k-1}{k}\,\frac{k+2}{12k-16}\,\left(\frac{2k+4}{6k-3}\right)^\frac{2k+4}{4k-7}\,,
\end{align}
where $g_{\rm *}$ reads the relativistic degrees of freedom at the point of maximum radiation temperature. With non-minimal contribution taken into account, we note that reheating can be completed before the onset of BBN for $k>4$ by properly adjusting the non-minimal coupling. In the left panel of Fig.~\ref{fig:TRH}, we show the maximum temperature during gravitational reheating for different choices of $\xi$, with $\xi=0$ being the minimal scenario. A larger $\xi$ results in low $\Tmax/\Trh$ for a given $k$, since the reheating temperature rises with $\xi$ (while $\Tmax$ stays approximately constant with $k$ for a given $\xi$), as evident from the right panel. The reheating temperature must satisfy the lower bound from the Big Bang nucleosynthesis (BBN), which requires  
$\Trh>T_{\rm BBN}\simeq 4$ MeV~\cite{Sarkar:1995dd, Kawasaki:2000en, Hannestad:2004px, DeBernardis:2008zz, deSalas:2015glj,Hasegawa:2019jsa}, in order not to spoil prediction from BBN. Note that lowering $k$ also results in low $\Trh$, since in that case, the inflaton energy density redshifts slower (see Eq.~\eqref{eq:rhoPh}) and the reheating is completed at a much later time. $\xi=0$ corresponds to the minimal scenario associated with the lowest possible reheating temperature in the gravitational reheating setup indicated along the boundary of the red-shaded region. Therefore, there is no reheating via gravitational interaction within the red-shaded region.  

\section{Flavoured leptogenesis during gravitational reheating}
\label{sec:lepto}
We now discuss the generation of baryon asymmetry via leptogenesis by taking into account the effect of equilibration of the charged lepton Yukawa interactions during gravitational reheating. The relevant interaction Lagrangian responsible for leptogenesis, is given by
\begin{align}
-\mathcal{L}=\overline{\ell}_{L_\alpha}\, (y_{\nu})_{\alpha i}\,\widetilde{H}\, N_{i}+ \frac{1}{2}\,\overline{N_{i}^c}\,\left(M\right)_{ii}\,N_i+ \text{h.c}.,~(i=1,\,2,\,3)\,,
 \end{align}
where the SM particle content is extended with the addition of three generations of right-handed neutrinos $N_i$, singlet under the SM gauge symmetry. The SM left-handed leptons doublets are identified as $\ell_{L_{\alpha=e,\mu,\tau}}$ and $\widetilde{H}=i\,\sigma_{\rm 2}\,H^{\rm *}$ where $H$ represents the SM Higgs doublet. $\sigma_{\rm i}$ are the Pauli spin matrices. We assume the Majorana masses $M_i$ to be hierarchical $M_1\ll M_{2,3}$. We also consider that lepton-number-violating interactions of $N_1$ are rapid enough to wash out the asymmetry originating from the decay of the other two. Therefore, only the CP-violating asymmetry from the decay of $N_1$ survives and is relevant for leptogenesis. In that case, the structure of the Yukawa coupling $y_{\nu}$ can be constructed using Casas-Ibarra (CI) parametrization~\cite{Casas:2001sr} as (see Appendix.~\ref{sec:CI} for details):
\begin{align}\label{eq:Ynu}
y_{\nu}=-i \frac{\sqrt{2}}{v} \mathcal{U} D_{\sqrt{m}} \mathcal{R}^T D_{\sqrt{M}},
\end{align}
where $\mathcal{U}$ is the Pontecorvo-Maki-Nakagawa-Sakat (PMNS) matrix \cite{Zyla:2020zbs}, which connects the flavour basis with the mass basis for light neutrinos. $D_{\sqrt{m}}=\text{diag}(\sqrt{m_1},\,\sqrt{m_2},\,\sqrt{m_3})$ is the diagonal matrix containing the square root of light neutrino mass and similarly $D_{\sqrt{M}}= \text{diag} (\sqrt{M_1},\,\sqrt{M_2},\,\sqrt{M_3})$ represent the diagonal matrix for the RHN masses. Here $\mathcal{R}$ is an orthogonal matrix satisfying $\mathcal{R}^{\rm{T}}\mathcal{R}=1$. The CP asymmetry $\epsilon_{\Delta L}$ is generated due to the interference between tree level and one loop level decay of the RHN to lepton(anti-lepton) and Higgs doublets and can be parameterized after performing the flavour sum as~\cite{Davidson:2008bu}
\begin{align}
    \epsilon_{\Delta L} = \sum_\alpha \epsilon _{1\alpha}^\ell = \frac{1}{8\pi (y_\nu^\dagger y_\nu)_{11}} \sum_{j\neq1} \text{Im} \left[(y_\nu^\dagger y_\nu)_{1j}^2)\right] \mathcal{F}\left(\frac{M_j^2}{M_1^2}\right)\,,
\end{align}
where 
$    \mathcal{F}(x) \equiv \sqrt{x}\,\left[\frac{1}{1-x}+1-(1+x)\,\log\left(\frac{1+x}{x}\right)\right]\,$ is the loop factor. However, depending on the equilibration temperatures of different charged lepton Yukawa interactions mediated by the Yukawa interaction $\ell_L H E_R$, $E_R$ being the right-handed leptons in the SM, CP asymmetry along individual lepton flavour direction may become important, leading to the different ``flavour" regimes of leptogenesis.
\begin{figure}[htb!]
    \centering    \includegraphics[scale=0.36]{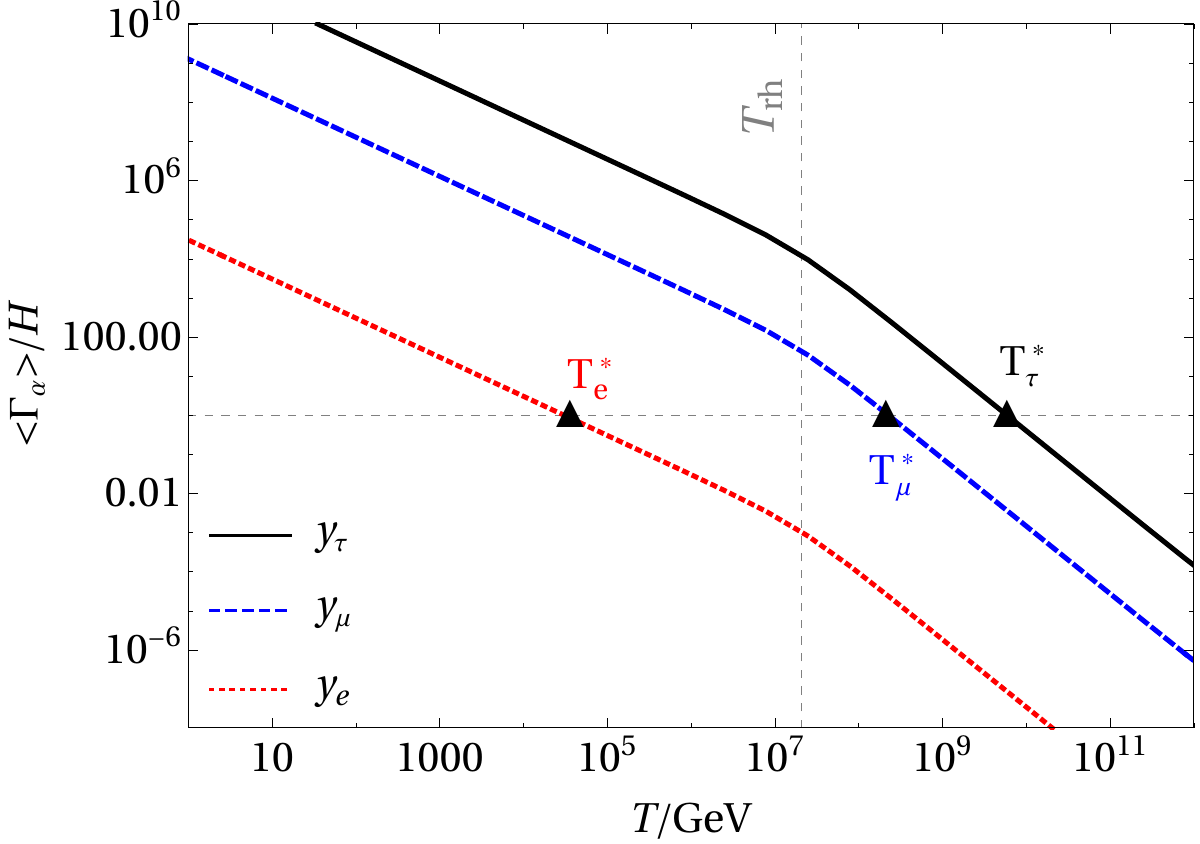}~~~\includegraphics[scale=0.36]{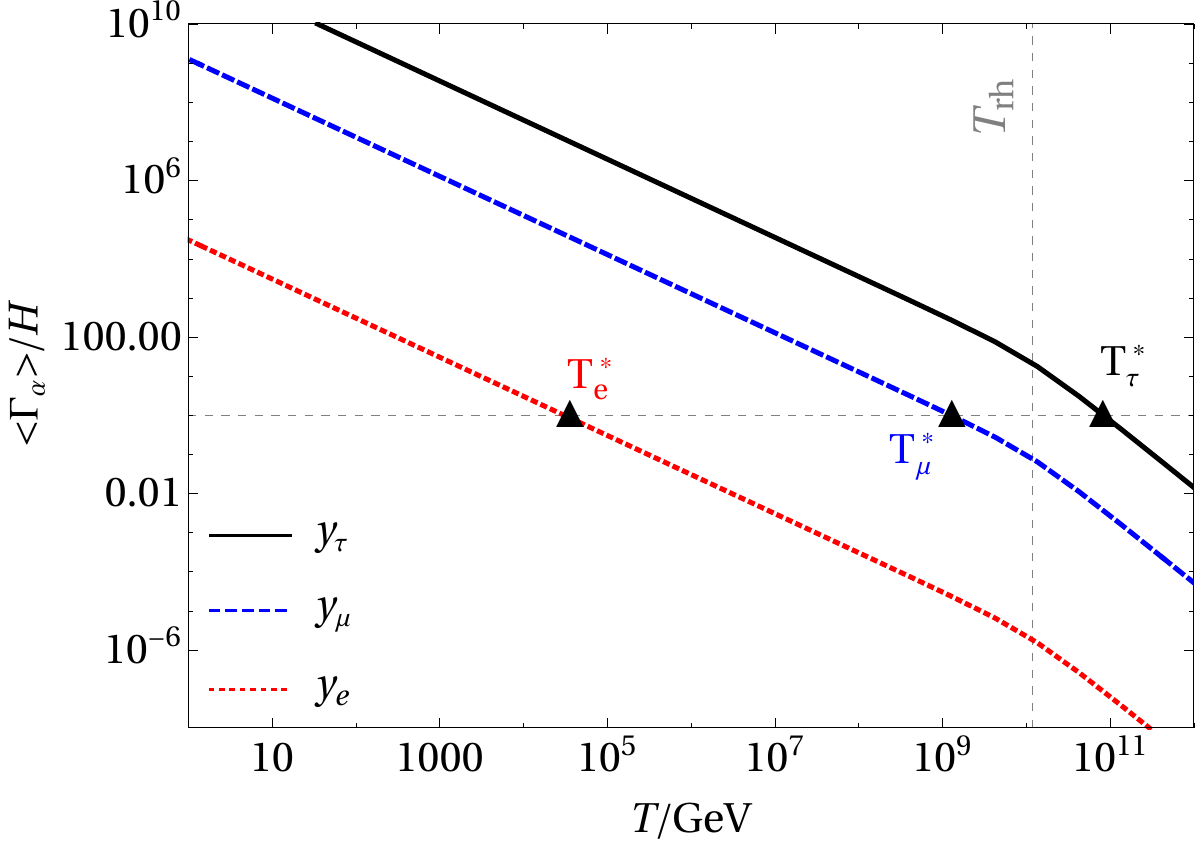}
    \caption{{\it Left:} Minimal gravitational reheating. We compare the ratio of reaction rate for charged lepton Yukawa interaction [Eq.~\eqref{eq:gamma1}] to the Hubble rate [Eq.~\eqref{eq:hub1}] as a function of the bath temperature. Different curves correspond to different flavours, as indicated. The triangular shapes show the point of equilibration of the corresponding flavour interaction. {\it Right:} Same as that of the left, but now for the non-minimal reheating scenario, with $\xi=5$. The vertical gray dashed line indicates the reheating temperature in each case. The horizontal gray dashed line indicates $\langle\Gamma_\alpha\rangle=H$. In all cases, we choose $k=20$.}
    \label{fig:rate}
\end{figure}
\begin{figure}[htb!]
    \centering
    \includegraphics[scale=0.36]{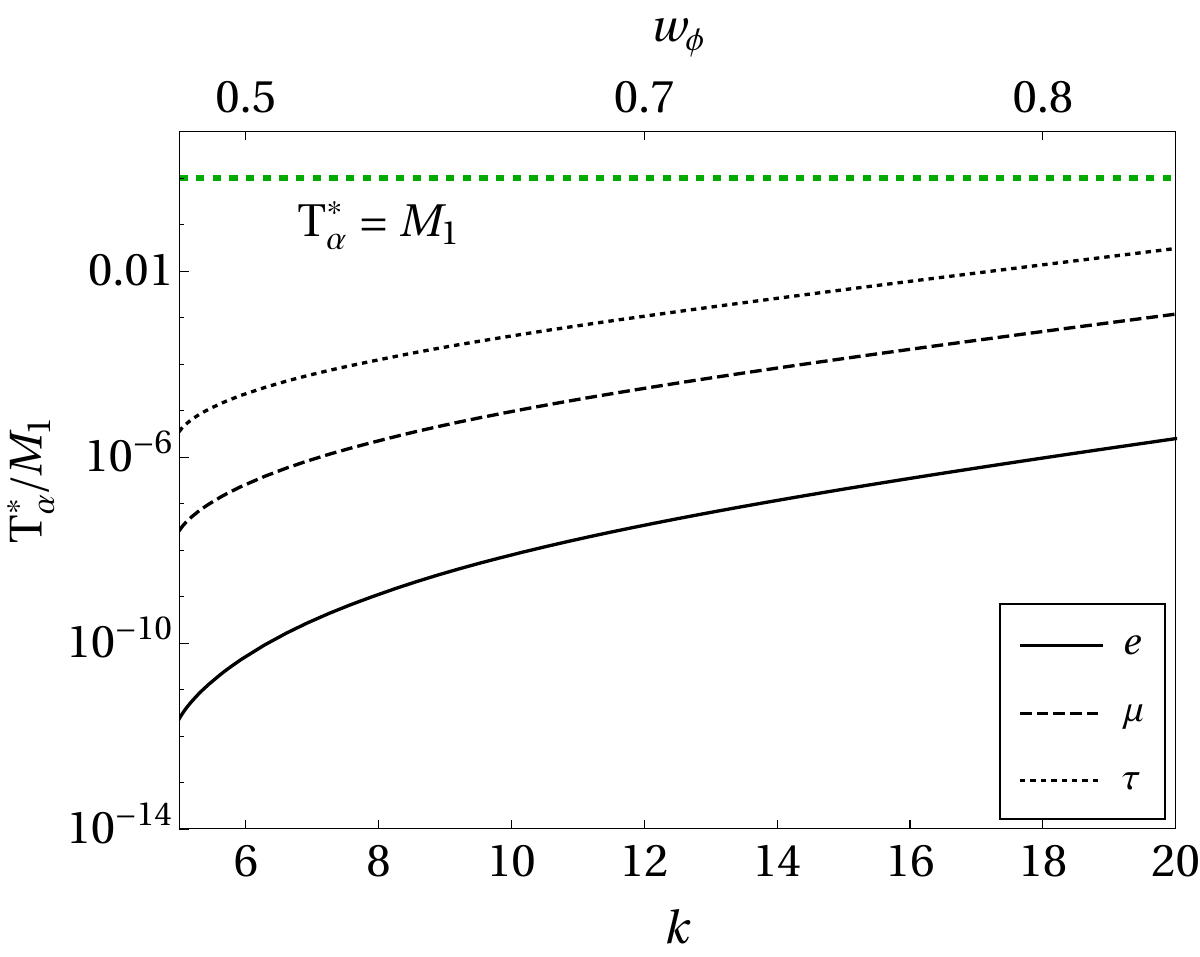}
   \includegraphics[scale=0.35]{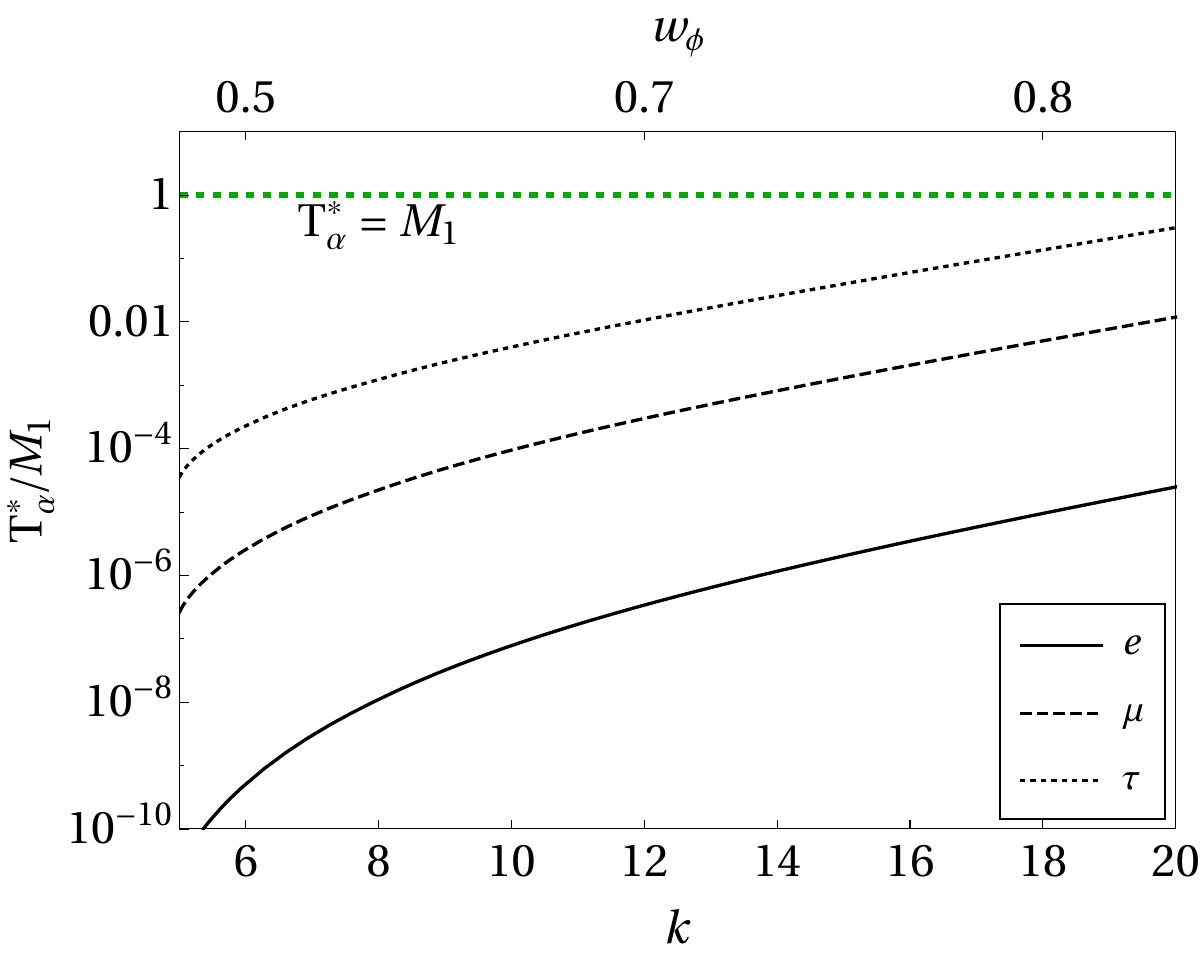}\\[10pt]
    \includegraphics[scale=0.45]{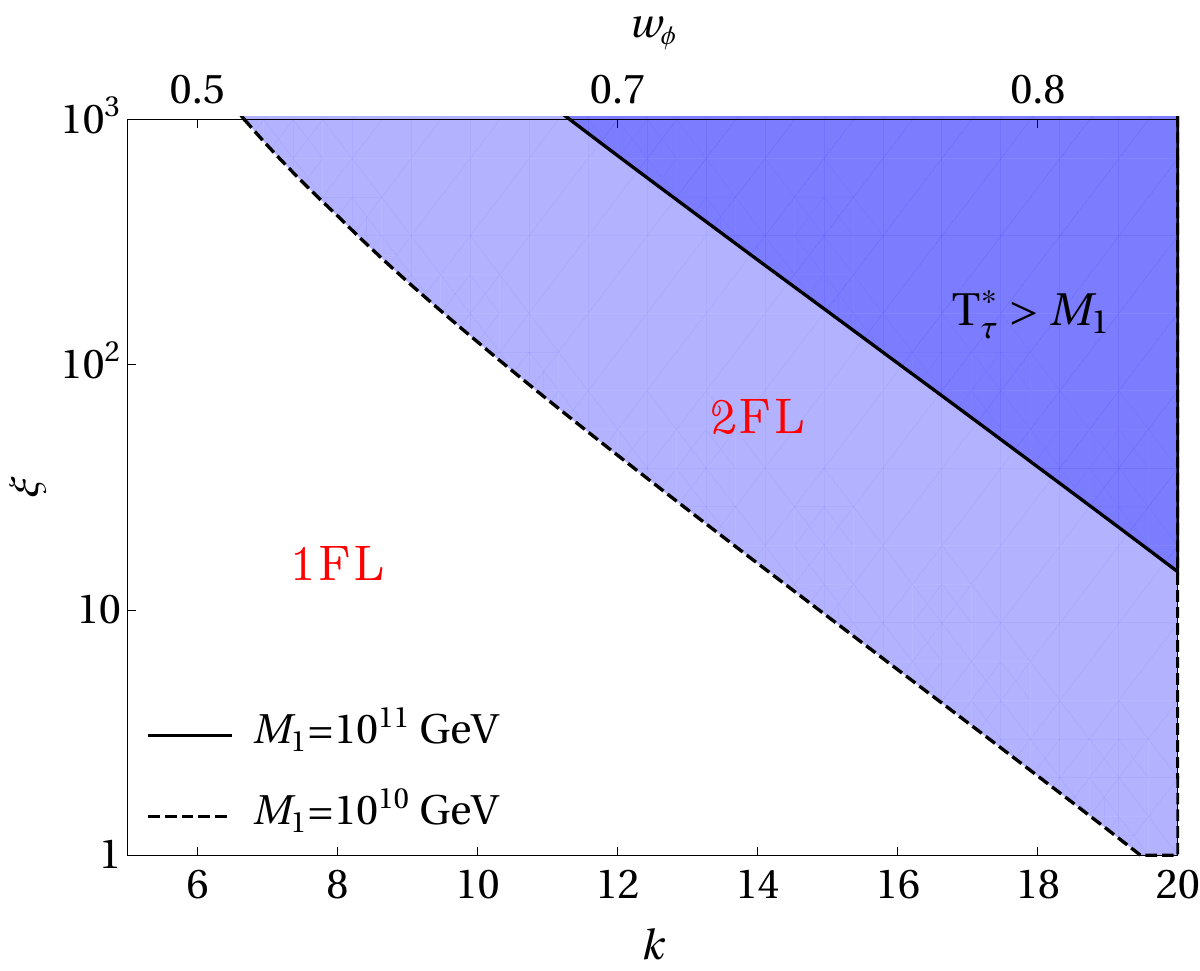}
    \caption{{\it Top:} Minimal gravitational reheating. We show the ratio of equilibrium temperature to RHN mass, as a function of $k$ in the, for different lepton flavours, with $M_1=10^{11}$ GeV (left) and $M_1=10^{10}$ GeV (right). {\it Bottom:} Non-minimal gravitational reheating. The solid (dashed) contours correspond to $T_\alpha^\star=10^{11}\,(10^{10})$ GeV. Within all shaded regions (marked as ``2FL"), one needs to consider two lepton flavours (see text).}
    \label{fig:Tst}
\end{figure}

To evaluate the equilibration temperature $T^*_\alpha$ of such processes, one needs to  compare the relevant interaction rates with the Hubble expansion rate of the Universe i.e. solve for the equation $\langle \Gamma_\alpha\rangle/H=1$, where for charged lepton Yukawa interactions of flavour $\alpha$, the relevant interaction rate (thermally averaged) can be approximated as~\cite{Datta:2022jic,Datta:2023pav}
\begin{align}\langle\Gamma_\alpha\rangle \simeq 5 \times 10^{-3}\,y_\alpha^2\,T,
    \label{eq:gamma1}
\end{align}
with $y_\alpha= \{2.94\times 10^{-6},\,6.04 \times 10^{-4},\,0.01\}$ representing the numerical estimates of Yukawa couplings for the charged lepton Yukawa interactions of $e,\,\mu,\,\tau$ flavour respectively in the charged lepton diagonal basis. During reheating, this leads to
\begin{align}\label{eq:Tst} 
T^{\star(\rm rh)}_{\alpha}\simeq\left(\frac{M_P}{5 \times 10^{-3} \sqrt3}\right)^{\frac{k+2}{2(1-k)}}\,\left(\frac{\pi^2 g_*}{30\sqrt3\alpha_k^\xi\,M_P^4}\right)^{\frac{3k}{8(1-k)}}\,\left(\frac{\rho_e}{4 M_P^4}\right)^{\frac{7-4k}{8(1-k)}}\,\left(\frac{8k-14}{k+2}\right)^{\frac{3k}{8(1-k)}}\,y_\alpha^{\frac{k+2}{k-1}}\,,
\end{align}
while for a standard radiation-dominated background,
\begin{align}\label{eq:TstRD}
&  T^{\star(\rm rd)}_{\alpha}\simeq \frac{15\times 10^{-3}}{\pi}\,y_\alpha^2\,\sqrt{\frac{10}{\gs}}\,M_P\,.   
\end{align}

In Fig.~\ref{fig:rate}, we show the epoch of equilibration of charged Yukawa interaction for the three lepton flavours by comparing the interaction rate (see Eq.~\eqref{eq:gamma1}), with the Hubble rate (see Eq.~\eqref{eq:hub1}) during reheating. For the minimal gravitational coupling, as shown in the left panel, both $\tau$ and $\mu$ interactions equilibrate during reheating. This can be verified from Eq.~\eqref{eq:non-minimal-trh}, which gives rise to $\Trh^{\xi=0}\simeq 2.1\times 10^7$ GeV with $k=20$, while from Eq.~\eqref{eq:Tst}, we find $\{T^\star_\mu,\,T^\star_\tau\}\simeq\{ 10^8,\,3.0\times 10^9\}$ GeV. For non-minimal scenario, with $\xi=5$, we find $\Trh^{\xi=5}\simeq 1.2\times 10^{10}$ GeV, whereas $\{T^\star_\mu,\,T^\star_\tau\}\simeq\{ 10^9,\,4.3\times 10^{10}\}$ GeV, implying, the Yukawa interaction corresponding to $y_\mu$ equilibrates during radiation domination. In this case, the equilibration temperature $T^\star_\mu$ is given by Eq.~\eqref{eq:TstRD}. In both cases, we find it is not possible for the electron Yukawa to achieve $\langle\Gamma_e\rangle>H$ during reheating.     

It can easily be realized from the top panels of Fig.~\ref{fig:Tst}, contrary to the standard radiation domination phase, charged lepton Yukawa interactions of all three flavours remain always in out-of-equilibrium during the time of lepton asymmetry generation around the temperature $T\simeq M_1$ with $M_1=\{10^{10},\,10^{11}\}$ GeV for the minimal gravitational reheating scenario. Consequently, so-called unflavoured (1FL) prescription would be sufficient to evaluate the exact amount of generated lepton asymmetry. This situation however changes when the non-minimal contribution to radiation energy density is introduced, as shown in the bottom panel for the same set of RHN masses. In that case, with the increased strength of the non-minimal coupling, it is possible to achieve $T^\star_\tau>M_1$ (denoted by the shaded blue region in the bottom panel)\footnote{Note that, the condition $T_{e,\mu}^*>M_1$ can not be obtained during reheating with our choice of RHN masses, as well as $k$-value because the maximum values of $T_{e,\mu}^*$ is attained at the time of RD, with $T_{e,\mu}^*<M_1$. Moreover, With prolonged gravitional (minimal or non-minimal) reheating phase, the value of $T_{e,\mu}^*$ would always be less compared to that in the standard RD case.}. For example, with $k=16$ and $M_1= 10^{11}$ GeV, $T^\star_\tau>M_1$ occurs for $\xi\gtrsim 10^2$ while for $M_1=10^{10}$ GeV, the limit on $\xi$ is sufficiently relaxed to $\xi\gtrsim 5$. Consequently as $\tau$-lepton Yukawa is in equilibrium during the time of leptogenesis at $T\sim M_1$, one, therefore, needs to take care of the lepton asymmetry generated along $\tau$ flavour direction and along $(e+\mu)$ flavour direction separately. This gives rise to the so-called two-flavour (2FL) regime, which is equally valid for the shaded region of the bottom panel of Fig.~\ref{fig:Tst}, whereas within the unshaded region, no lepton Yukawa inetraction equilibrates, resulting in an unflavoured or 1FL regime. 

To learn about the generation and subsequent evolution of the lepton (more specifically $B-L$) asymmetry, we need to solve the following set of coupled BEQs along with Eq.~\eqref{eq:BErhop} and \eqref{eq:BErhoR} that track the evolution of RHN number density, as well as the $B-L$ asymmetry,
\begin{align}\label{eq:BEQ} 
& \Dot{n}_{N_1}+ 3\,H\,n_{N_1}= R_{N}^\phi - \langle \Gamma_N\rangle\,\left(n_{N_1}-n_{N_1}^{\rm eq}\right)+R_{N}^T
\nonumber\\&
\Dot{n}_{B-L}+ 3\,H\,n_{B-L}=-\langle \Gamma_N\rangle\,\left[\epsilon_{\Delta L}\,(n_{N_1}-n_{N_1}^{\rm eq})+\frac{n_{N_1}^{\rm eq}}{2\,n_\ell^{\rm eq}}\,n_{B-L}\right]
\nonumber\\&
3\,H^2\,M_P^2= \rho_\phi+\rho_R+ n_{N_1}\,E_N
\nonumber\\&
E_N^2=M_1^2 + 9 T^2\,,
\end{align}
where
\begin{align}
\langle \Gamma_N\rangle= \frac{K_1(M_1/T)}{K_2(M_1/T)}\frac{M_1}{8\pi} \left(y_\nu^\dagger y_\nu\right)_{11}\,,    
\end{align}
is the thermally averaged RHN decay width and $K_i[z]$ represents the modified Bessel function. We also define the RHN yield as $Y_N=n_{N_1}/s$, where $s=\left(2\pi^2/45\right)\,\gss(T)\,T^3$ is the entropy per comoving volume, with $\gss(T)$ being the number of relativistic degrees of freedom associate with entropy at temperature $T$. Here
\begin{align}
    n_j^{\rm eq}= \frac{g_j\,T^3}{2 \pi^2} \left(\frac{M_j}{T}\right)^2 K_2(M_j/T),
\end{align}
is the equilibrium number density of a species $j$,
with $g_j$ being the corresponding degrees of freedom. 

The {\it minimal} RHN production rate density from scattering of a pair of inflaton condensate mediated by graviton (shown in Fig.~\ref{fig:feyn}) reads~\cite{Clery:2021bwz,Clery:2022wib,Barman:2022qgt}
\begin{align}\label{eq:RphN}
& R_{N}^\phi=\frac{\rho_\phi^2}{4\pi M_P^4}\,\frac{M_i^2}{m_\phi^2}\,\Sigma_{N_i}^k\,,   
\end{align}
where
\begin{align}
\Sigma_{N}^k = \sum_{n=1}^{+\infty} |{\cal P}_{2n}^k|^2\,\frac{m_\phi^2}{E_{2n}^2}\,
\left[1-\frac{4\,M_{N}^2}{E_{2n}^2}\right]^{3/2}\,,    
\end{align}
accounts for the sum over the Fourier modes of the inflaton potential. Here $E_n = n \omega$ is the energy of the $n$-th inflaton oscillation mode. Similarly, the {\it minimal} RHN production rate density from the SM bath, mediated by graviton, is given by~\cite{Mambrini:2021zpp,Bernal:2021kaj,Barman:2021ugy,Clery:2021bwz,Haque:2021mab,Clery:2022wib,Barman:2022qgt}
\begin{align}\label{eq:RTN}
& R_{N}^T\simeq 1.7\times 10^{-2}\,\frac{T^8}{M_P^4}\,.  
\end{align}
Note that there will be another contribution to the RHNs from  $\langle \Gamma_N\rangle\,\left(n_{N_1}-n_{N_1}^{\rm eq}\right)\times E_N$ due to the inverse decay from the thermal bath (see, for instance, Eq.~\eqref{eq:rhoR}) \footnote{It is important to note here, in the presence of the RHNs, will have another contribution due to the decay of the RHNs into the SM final states, providing a source term to the radiation. However, such a contribution has a sub-dominant effect on radiation energy density.}. From Eq.~\eqref{eq:RphN} and \eqref{eq:RTN} we note,
\begin{align}
& \frac{R^\phi_{N}}{R^T_{N}}\Bigg|_{T=M_{N}}\approx\left(\frac{\rp}{m_\phi\,M_{N}^3}\right)^2\,, 
\end{align}
which shows, for the minimal scenario, RHN production shall always be dominated by inflaton condensate scattering over scattering of the bath particles, since $\rho_\phi\gg m_\phi\,M_N^3$. Furthermore, because of the non-minimal coupling, we will have a non-minimal contribution to RHN production on top of the minimal productions given by Eqs.~\eqref{eq:RphN} and \eqref{eq:RTN}. However such contributions are sub-dominant unless the $\xi$-values are very large (see Appendix.~\ref{sec:non-minimal} for details). As such large values of $\xi$ are not of interest in studying the flavour effects in the present framework, it is therefore sufficient to consider minimal production for the RHNs (and hence the asymmetry), with non-minimal contribution to radiation.   

In order to solve Eq.~\eqref{eq:BEQ}, we make the following re-definition of variables
\begin{align}
& \Phi=\rho_\phi\,A^{3\,(1+w_\phi)},\,R=\rho_R\,A^4,\,
N_1=n_{N_1}\,A^3,\,\,N_{B-L}=n_{B-L}\,A^3\,.
\end{align}
With such redefined variables the BEQs  takes the form
\begin{align}
& \Phi' = -(1+w_\phi)\,\frac{\Phi\,\Gamma_\phi}{A\,H}
\nonumber\\&
R'=(1+w_\phi)\,\frac{\Phi\,\Gamma_\phi}{A^{3w_\phi}\,H}+\frac{\langle\Gamma_N\rangle}{H}\,\left(N_1-N_1^{\rm eq}\right)\,E_N
\nonumber\\&
N_1' = \left(R_{N}^\phi+R_{N}^T\right)\,\frac{A^2}{H}-\frac{\langle\Gamma_N\rangle}{A\,H}\,\left(N_1-N_1^{\rm eq}\right)
\nonumber\\&
N_{B-L}' = -\frac{\langle\Gamma_N\rangle}{A\,H}\,\left[\epsilon_{\Delta L}\,(N_1-N_1^{\rm eq})+\frac{N_1^{\rm eq}}{2\,N_\ell^{\rm eq}}\,N_{B-L}\right]
\nonumber\\&
3\,H^2\,M_P^2 = \frac{1}{A^3}\,\left(\frac{\Phi}{A^{3w_\phi}}+\frac{R}{A}+N_1\,E_N\right)
\nonumber\\&
E_N = \sqrt{M_1^2+9\,T^2}\,.
\end{align}
Here primes denote derivative with respect to $A=a/\aend$ while 
the Hubble parameter in Eq.~\eqref{eq:hub1} has been modified with contribution from RHN energy density. The $B-L$ asymmetry generated in this way gets converted to an effective baryon asymmetry due to the sphaleron transition around the sphaleron decoupling temperature $T_{\rm sph}\simeq 150$ GeV via~\cite{Harvey:1990qw} 
\begin{align}
Y_B\simeq c_{\rm sph}\,Y_{B-L}=c_{\rm sph}\,\frac{n_{B-L}}{s}=\frac{8\,N_F+4\,N_H}{22\,N_F+13\,N_H}\,Y_{B-L}\,,   
\end{align}
where $N_F$ is the number of fermion generations and $N_H$ is the number of Higgs doublets, which in our case: $N_F = 3,\,N_H = 1$, leading to $c_{\rm sph}\simeq 28/79$.

To this end we have adopted a flavor-blind approach, considering the fact that the lepton asymmetry is generated with no preferred flavor direction. Now, for $T_\alpha^\star>M_1$, one needs to look into the production and evolution of lepton asymmetry along that particular lepton flavor direction independently. In this case, the BEQ is modified to,
\begin{align}
    \Dot{n}_{\Delta_\alpha}+3 H n_{\Delta_\alpha}=- \langle\Gamma_{N} \rangle \left[ \epsilon^\ell_{1\alpha}(n_{N_1}-n_{N_1}^{\rm eq}) +K^0_\alpha \sum_\beta (C^\ell_{\alpha \beta} +C^H_{\alpha \beta})\frac{n_{N_1}^{\rm eq} n_{\Delta_\beta}}{n_\ell^{eq}}\right]\,,\Delta_{\alpha}=\frac{B}{3}-L_\alpha,
    \label{eq:bl-flavor}
\end{align}
with $\alpha$ encoding a particular flavor direction. For example, in case $T_\tau^\star>M_1$, which we realized in the bottom panel of Fig.~\ref{fig:Tst}, $\alpha\equiv\{a (=e+\mu),\,\tau\}$, implying $\tau$-mediated interactions are dominant compared to Hubble and thus distinguishable compared to the other two flavors. Consequently, $\epsilon^l_{1\alpha}$ would be the CP asymmetry generated along the $\alpha$ flavor and 
\begin{align}
K^0_\alpha=\frac{\Gamma(N_1\to \ell_\alpha+ H)}{\sum_\alpha\Gamma(N_1\to \ell_\alpha+ H)}\,,    
\end{align}
represents the ratio of the decay amplitudes at the tree level. The matrices $C^l$ and $C^H$ convert the asymmetry generated in $\ell_\alpha$ and $H$ states to $\Delta_{\alpha}$ direction as
 \begin{align}
&  \frac{n_{\ell_\alpha}-n_{\bar{\ell}_\alpha}}{n_{\ell_\alpha^{\rm eq}}}
     =C^\ell_{\alpha \beta}  \frac{n_{\Delta_\beta}}{n_\ell^{\rm eq}}, & 
     \frac{n_{H}-n_{H^\dagger}}{n_{H^{\rm eq}}}
     =C^H_{\beta}  \frac{n_{\Delta_\beta}}{n_\ell^{\rm eq}}.
 \end{align}
The structure of these $C^l$ and $C^H$ matrices depends on the flavor regimes as well as on the chemical equilibrium conditions for all the SM interactions that are in equilibrium during the generation of lepton asymmetry. For example, in the two-flavor scenario, the matrices $C^\ell$ and $C^H$  read~\cite{Nardi:2006fx},
 \begin{align}
     C^\ell=\frac{1}{460}\begin{pmatrix}
         196 & -24\\
         -9 & 156
     \end{pmatrix},
     ~
     C^H=\frac{1}{230} (41~~~~56). 
 \end{align}
Solving Eq.~\eqref{eq:bl-flavor} (instead of second equation of Eq.~\eqref{eq:BEQ}) along with relevant BEQs for $\rho_\phi$, $n_{N_1}$ we can determine the final asymmetry along $B/3-L_\alpha$ direction, which then converted to baryon asymmetry via
\begin{align}
Y_B=\frac{28}{79}Y_{B-L}=\frac{28}{79} \sum_\alpha Y_{B/3-L_\alpha}\,.
\end{align}
\begin{figure}[htb!]
\centering
\includegraphics[scale=0.2]{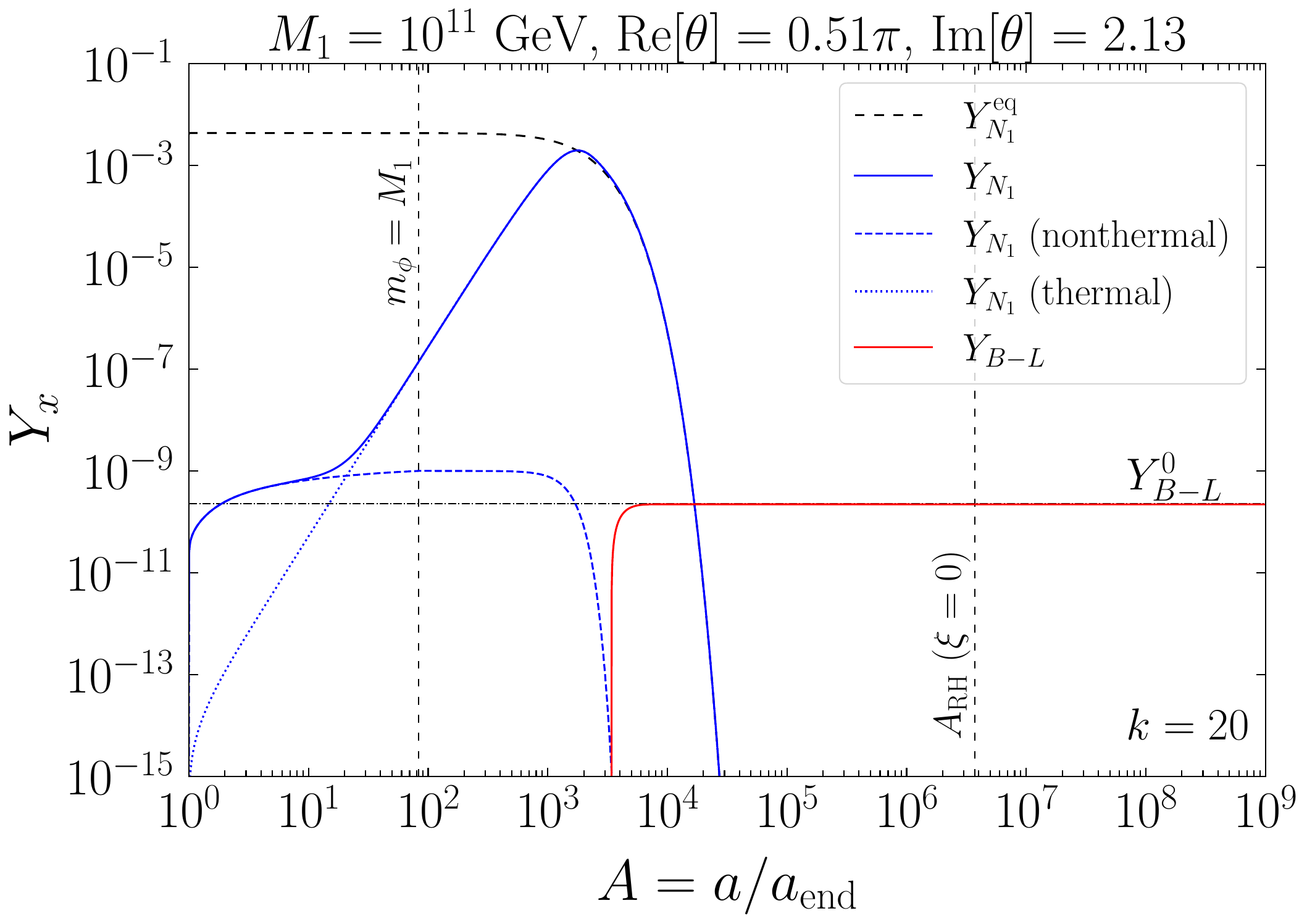}
\includegraphics[scale=0.2]{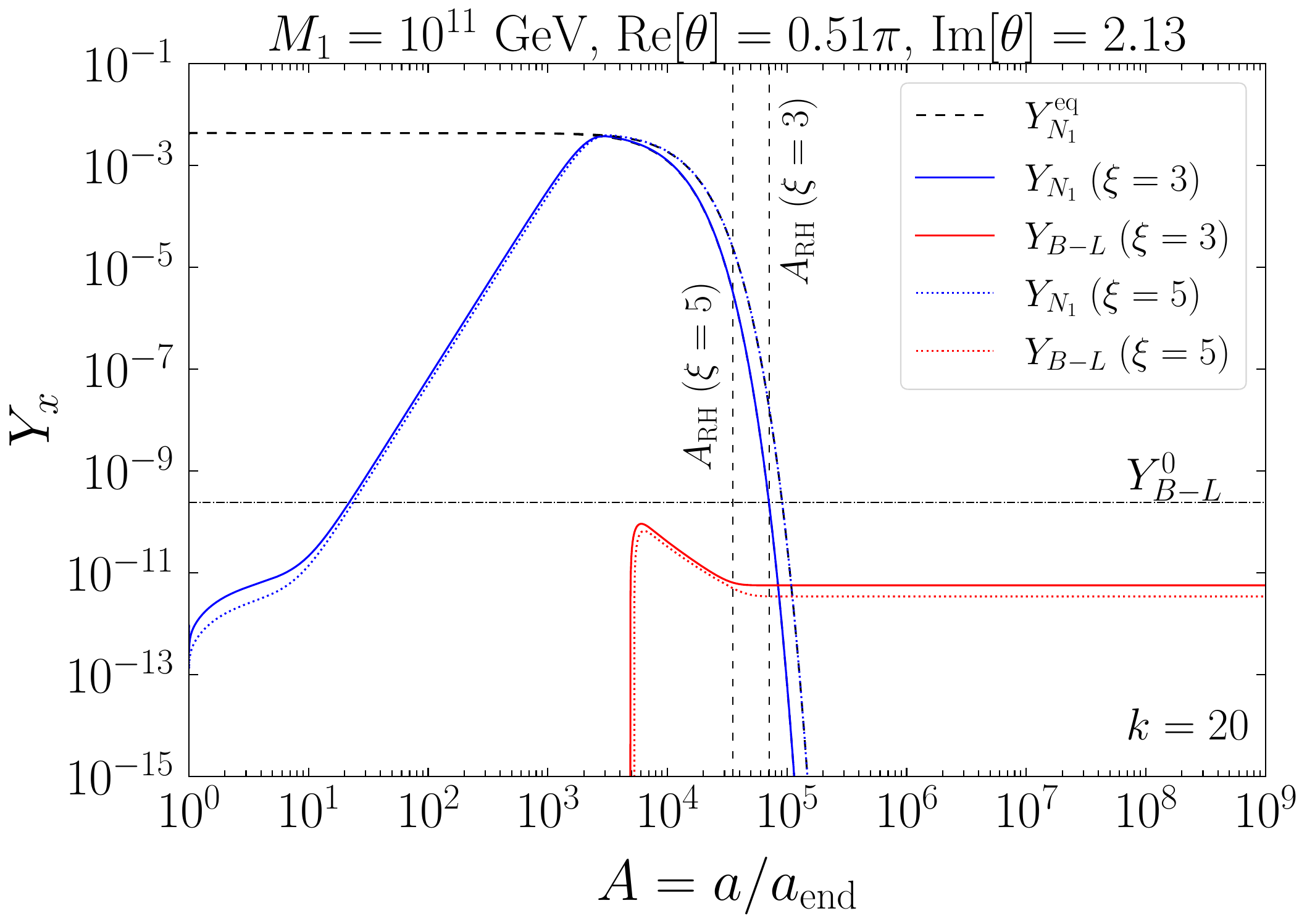}
\\[10pt]
    \includegraphics[scale=0.25]{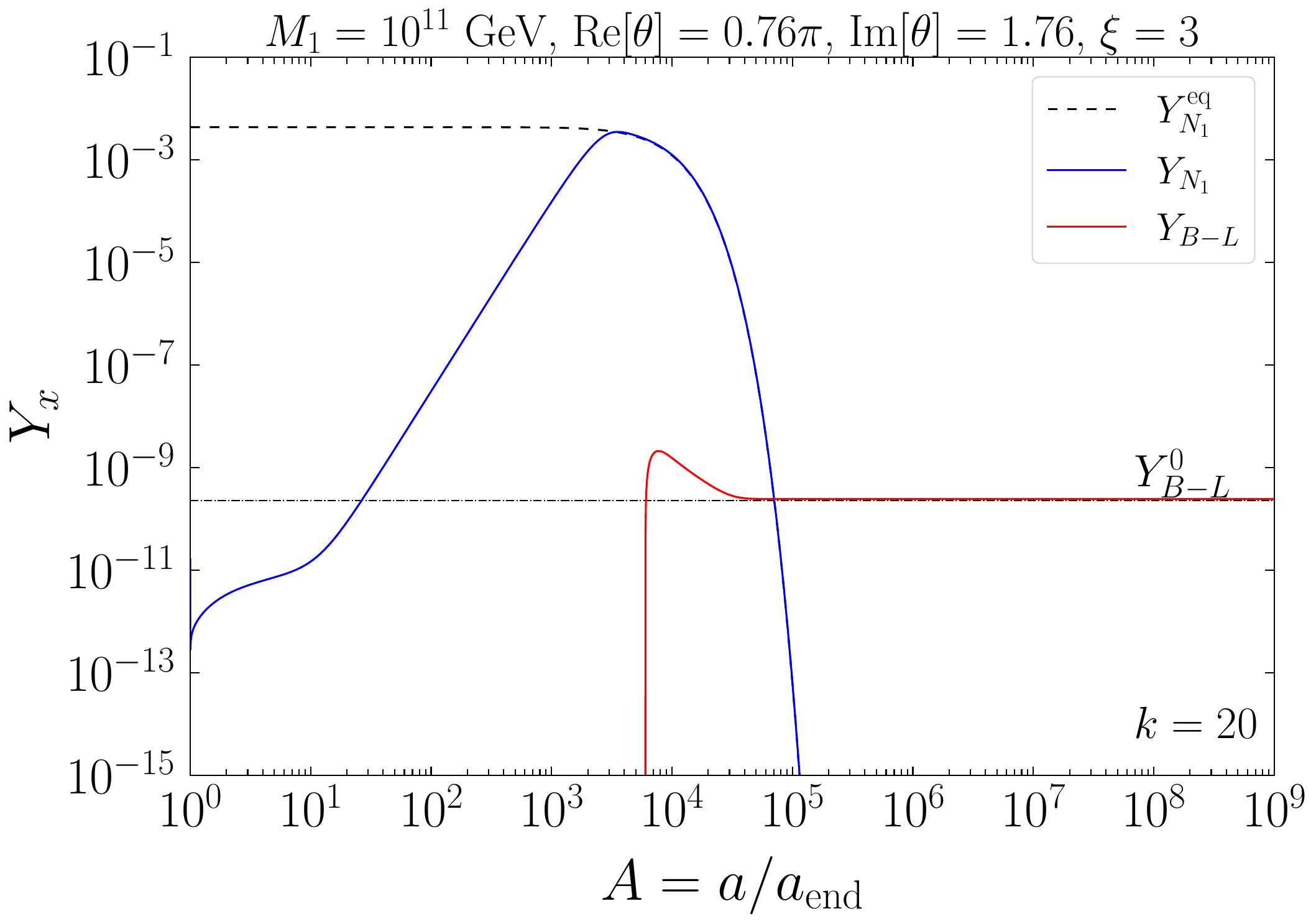}
\caption{{\it Top Left:} Minimal gravitational reheating. Evolution of RHN and $B-L$ (shown via red solid curve). Here we show individual contributions from bath (thermal) and gravity-mediated processes (non-thermal) via dotted and solid blue curves, respectively. {\it Top Right:} Non-minimal coupling. The solid and dotted curves correspond to $\xi=3$ and $\xi=5$, respectively. 
We fix $\text{Re}[\theta]=0.51\,\pi$ and $\text{Im}[\theta]=2.13$. {\it Bottom:} Same as top right, but now for a fixed $\xi=3$, with $\text{Re}[\theta]=0.76\,\pi$ and $\text{Im}[\theta]=1.76$, such that the observed baryon asymmetry is satisfied. For all cases we choose RHN mass $M_1=10^{11}$ GeV with $k=20$.
}
\label{fig:yield-min1}
\end{figure}
\begin{figure}[htb!]
    \centering
    \includegraphics[scale=0.2]{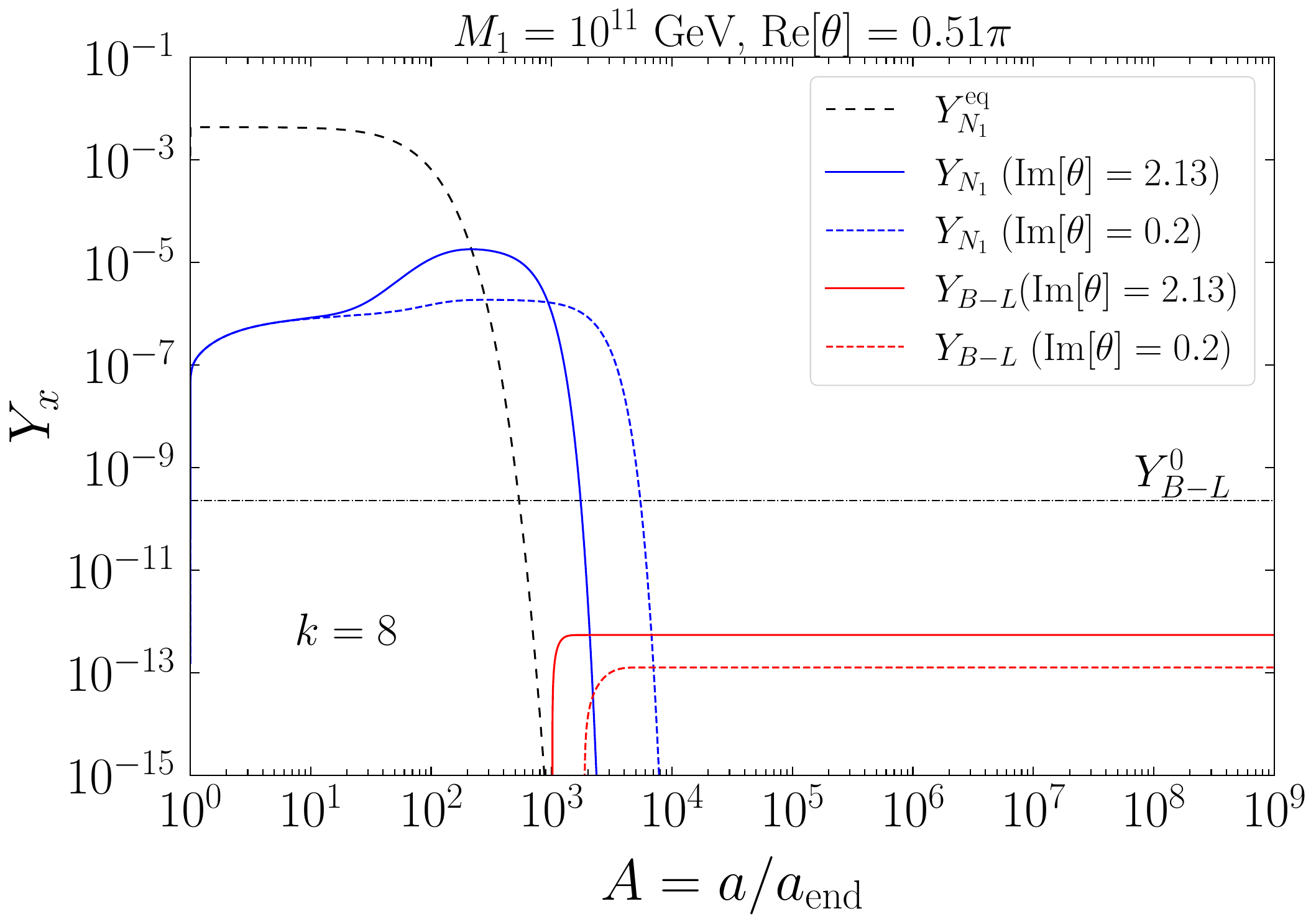}~\includegraphics[scale=0.2]{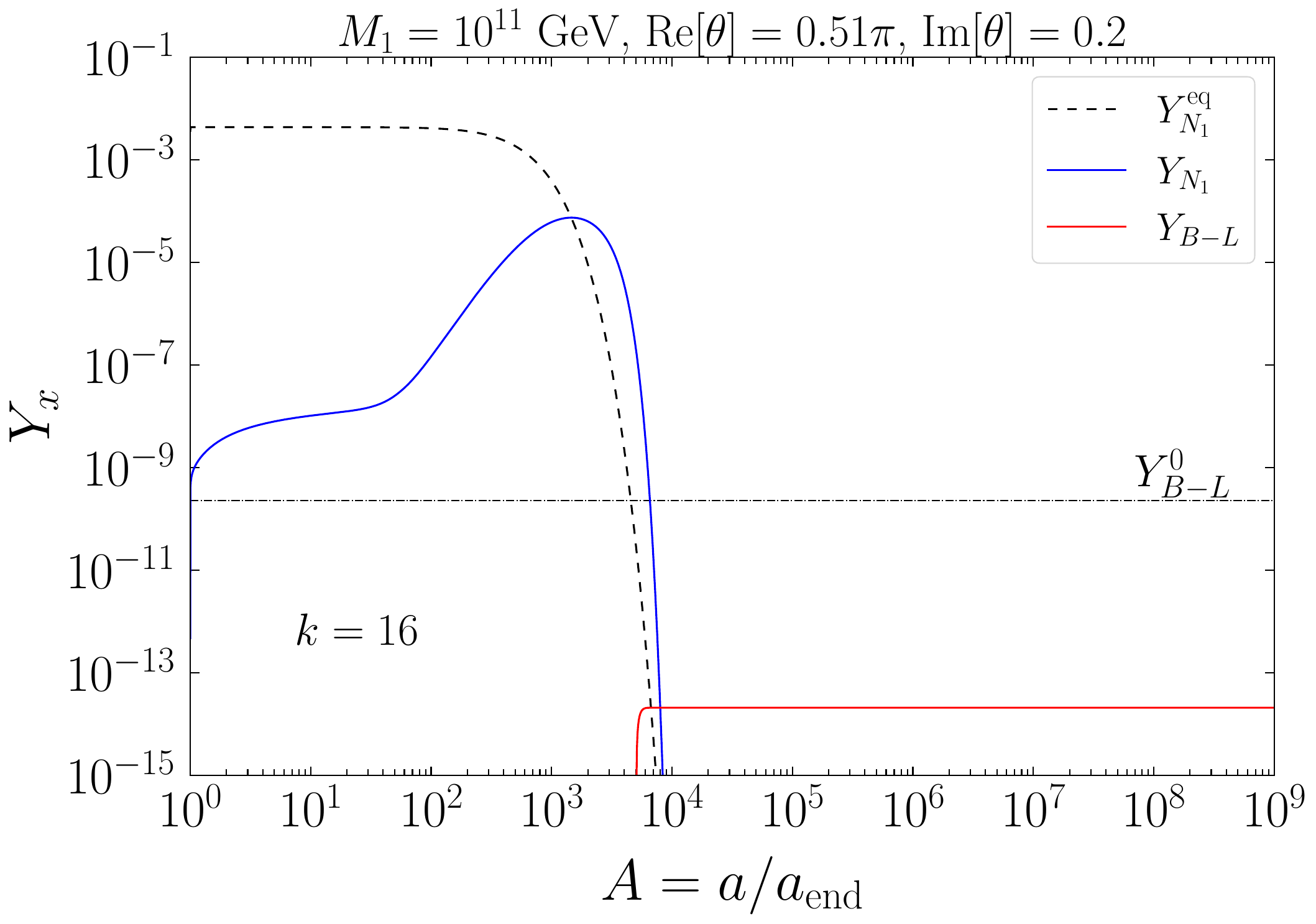}
    \caption{Minimal gravitational reheating. {\it Left:} Evolution of RHN and $B-L$ (shown via red solid curve) yield for $k=8$, where the solid and dashed curves correspond to $\text{Im}[\theta]=2.13$ and $\text{Im}[\theta]=0.2$, respectively, keeping $\text{Re}[\theta]=0.51\,\pi$. We also show yield from equilibrium distribution via black dotted curve. The black dashed horizontal straight line corresponds to observed baryon asymmetry. {\it Right:} same as left left but for $k=16$, with $\text{Re}[\theta]=0.7\,\pi$ and $\text{Im}[\theta]=0.2$. In all cases we fix $M_1=10^{11}$ GeV.}
    \label{fig:yield-min2}
\end{figure}
\begin{figure}[htb!]
    \centering
    \includegraphics[scale=0.32]{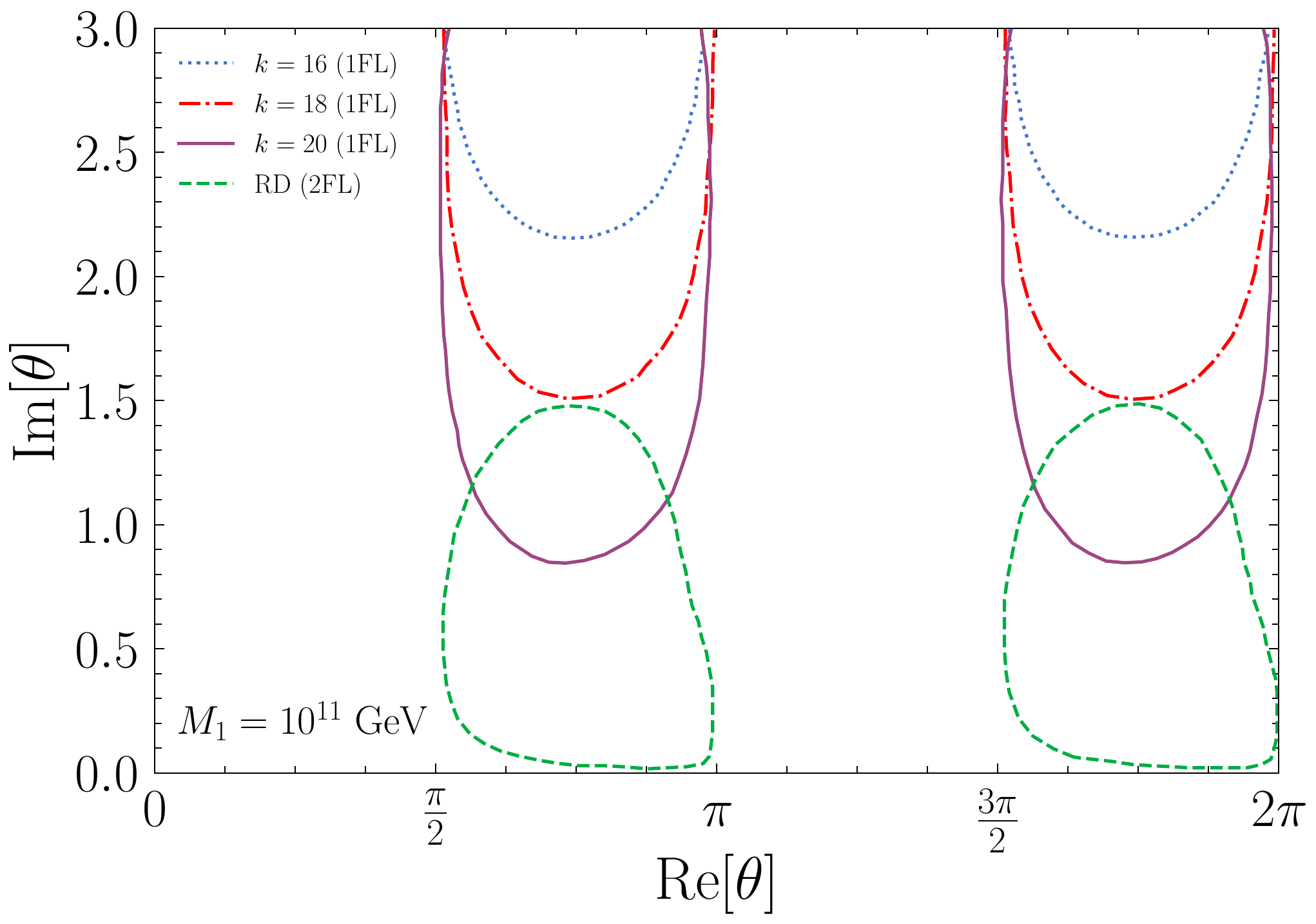}\\[10pt]
    \includegraphics[scale=0.32]{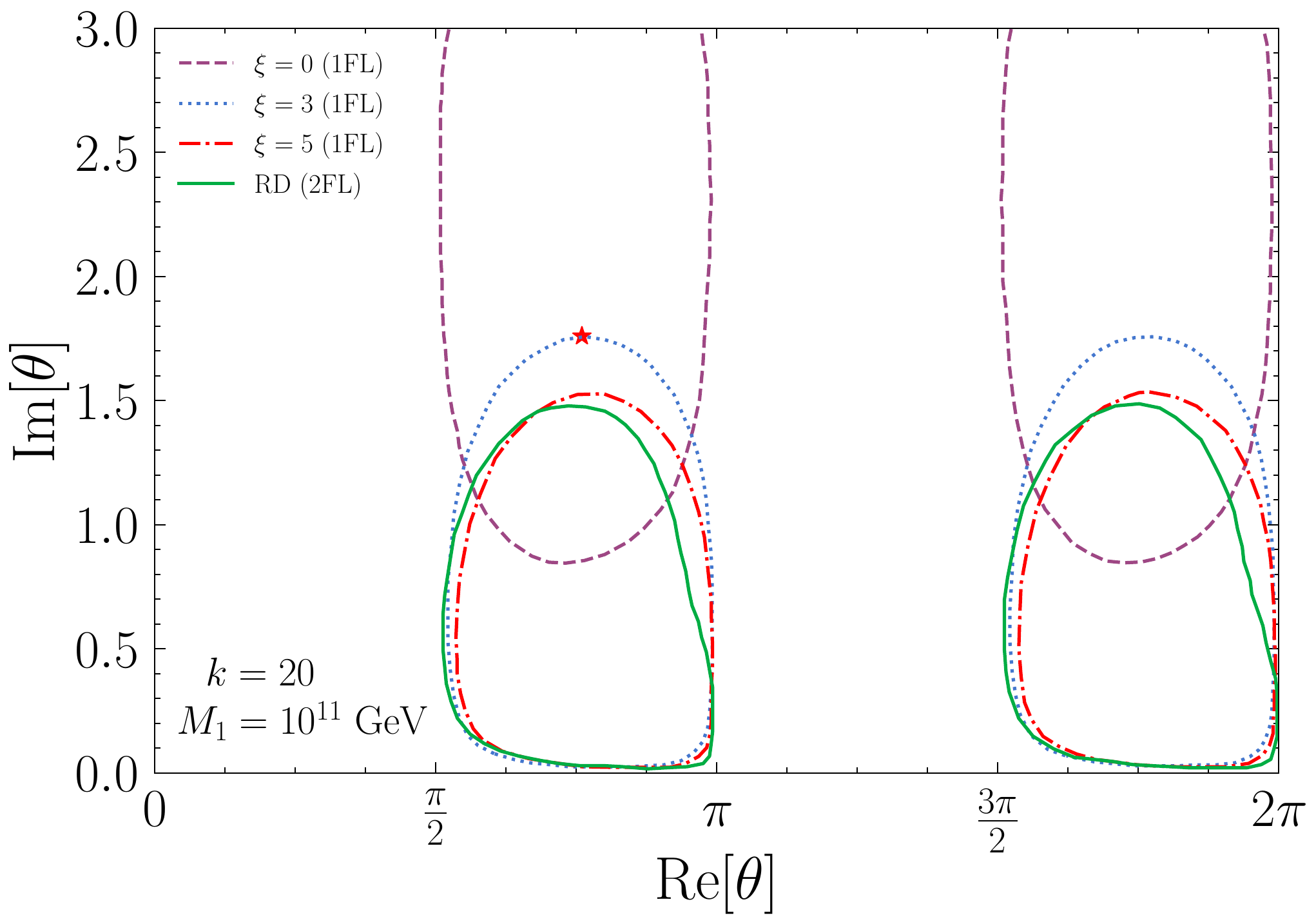}
    \caption{{\it Top:} Minimal scenario. All contours satisfy observed baryon asymmetry. The green dashed contours correspond to asymmetry production in a standard radiation-dominated Universe, while others correspond to asymmetry production during reheating for different choices of $k$ shown via different patterns. {\it Bottom:} Same as top, but for non-minimal scenario. The solid green contour corresponds to asymmetry production in a standard radiation-dominated Universe, while the other contours correspond to different choices of $\xi$ that produce the right asymmetry. The red star corresponds to the benchmark point in the bottom panel of Fig.~\ref{fig:yield-min1}. The RHN mass is fixed at $M_1=10^{11}$ GeV in all cases.}
    \label{fig:scan}
\end{figure}
\begin{figure}[htb!]
    \centering    \includegraphics[scale=0.32]{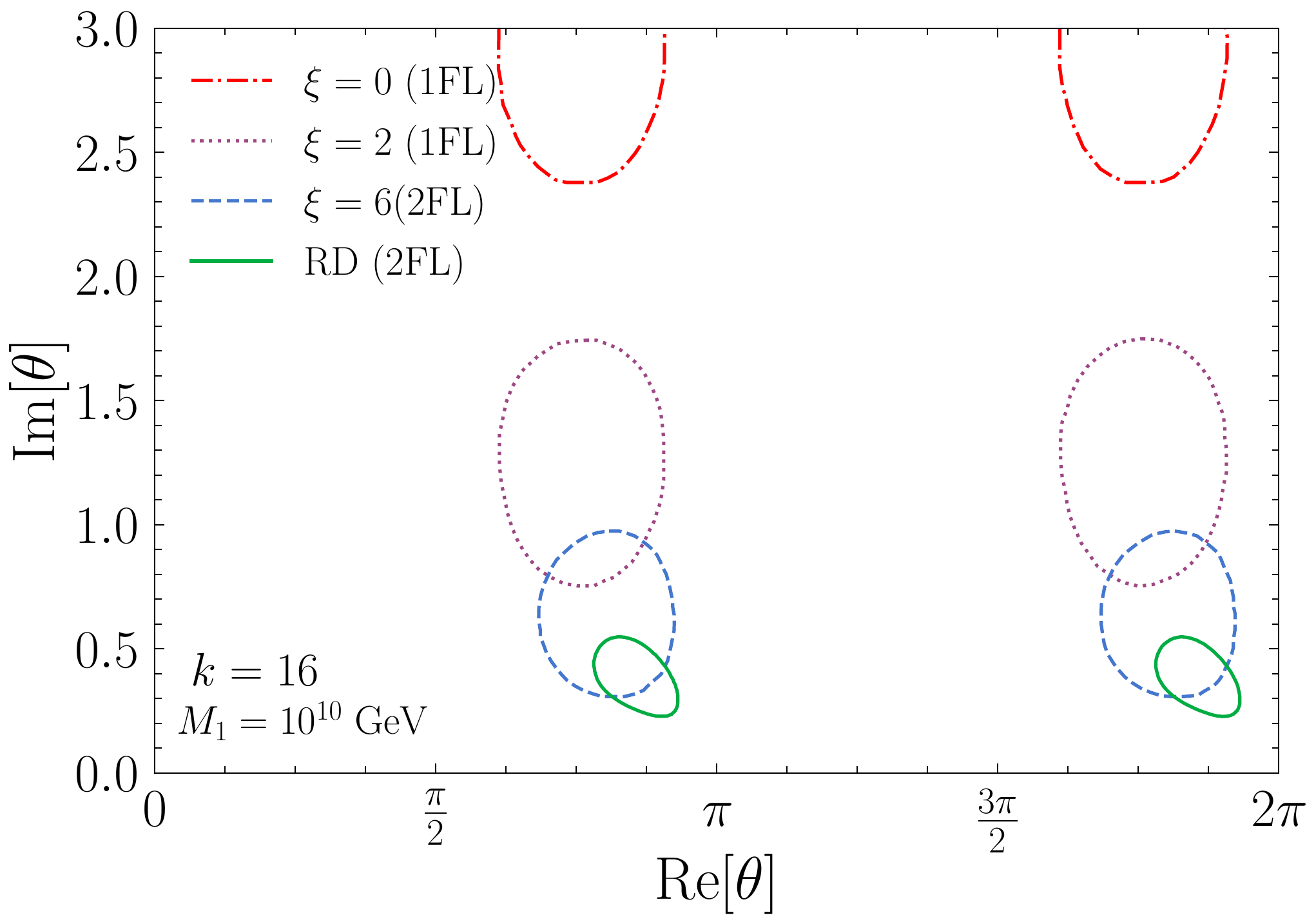}
    \caption{Same as Fig.~\ref{fig:scan}, but here for $M_1=10^{10}$ GeV. Different contours correspond to different choices of the non-minimal coupling $\xi$ that provides observed baryon asymmetry. Here 1(2)FL corresponds to 1(2) flavour regime.}
    \label{fig:scan2}
\end{figure}

In the top left panel of Fig.~\ref{fig:yield-min1}, we show the evolution of yield of RHN (solid blue curve) and $B-L$ asymmetry (red curve) for $M_1=10^{11}$ GeV, which eventually generates the correct amount of baryon asymmetry $Y_B=8.7 \times 10^{-11}$~\cite{Planck:2018jri}. Here we have two populations of RHNs: one coming from bath (denoted by blue dotted curve) and the other from inflaton scattering (indicated by blue dashed curve). As can be seen from the figure, initially just after the end of inflation, inflaton scattering process dominantly produces the RHNs. However, since the field-dependent inflaton mass decreases with time (see Eq.~\eqref{eq:inf-mass}) for $k>2$, there exists a point $a=a_c^\phi$ denoted by
\begin{align}
A_c^\phi\equiv\frac{a_c^\phi}{\aend} =\left(\frac{M_1}{m_{\rm end}}\right)^\frac{k+2}{3\,(2-k)}\implies A_c^\phi(k=20)\simeq 15\,\left(\frac{M_1}{10^{11}\,\text{GeV}}\right)^{11/17}\,,
\end{align}
beyond which the gravity-mediated production of RHNs from inflaton becomes kinematically forbidden\footnote{Note that the expression is not valid for $k=2$, as in that case, the inflaton mass is constant with $\mend\simeq 10^{13}$ GeV, and gravitational production from inflaton scattering happens throughout reheating as long as $m_\phi>M_N$.}. As a result, for $A> A_c^{\phi}$, inverse decay process takes over the scattering process for the RHN production till the effective decay of the RHNs starts producing the $B-L$ asymmetry. It is worth noting that the yield actually saturates even before $A=A_c^\phi$. This is attributed to the fact that, the inflaton energy density $\rho_\phi\propto a^{-\frac{6k}{k+2}}$ (see Eq.~\eqref{eq:rhoPh}), redshifts faster compared to the inflaton mass $m_\phi\propto a^{\frac{6-3k}{k+2}}$ (see Eq.~\eqref{eq:inf-mass}). Since $R_N^\phi\propto \rho_\phi^2$, hence the gravitational yield is stopped before the kinematical limit $m_\phi=M_1$ is achieved. 

The parameter dependence of asymmetry production can be understood from Fig.~\ref{fig:yield-min2}, wherein the left panel, we choose $k=8$ and show the evolution of the RHN yield, $B-L$ yield, together with the equilibrium yield, for two different choices of $\text{Im}[\theta]$ and for $M_1=10^{11}$ GeV. As can be seen, a larger $\text{Im}[\theta]$ (and therefore a larger $y_\nu$) results in a larger thermal contribution from the SM bath via $\ell_L + H\to N$ (or $\bar{\ell}_L + H^\dagger\to N$) to the RHN yield, and hence the corresponding larger asymmetry. On the other hand, a larger $k$ also generates a larger thermal contribution to RHN yield (as evident from the right panel of Fig.~\ref{fig:yield-min2}). This is expected because a larger $k$ results in larger $\Trh$, see Eq.~\eqref{eq:grav-trh}, and hence higher $\Tmax$, making it more feasible for the RHN to be produced from the thermal bath. However, 
since larger $k$ makes thermal contribution dominating over inflaton scattering, the RHN yield tends to track the equilibrium distribution during the decay of the RHN. As a consequence, for a few parameters, less amount of lepton asymmetry can be generated as seen by comparing the left and the right panel of Fig.~\ref{fig:yield-min2}.

Let us now discuss the effect of non-minimal contribution to the production of the bath on the lepton asymmetry generation. In the {\color{red}} right panel of Fig.~\ref{fig:yield-min1}, we show the evolution of yields with scale factor for different choices of $\xi=\{3,\,5\}$ (indicating the non-minimal effect). Other parameters are fixed at $M_1=10^{11}$ GeV and $k=20$. As we compare the $B-L$ yield (shown via red curves) we find that a larger $\xi$ results in the larger dilution of final asymmetry. This is attributed to the fact that a larger $\xi$ results in higher $\Trh$ (see Fig.~\ref{fig:TRH}), which eventually leads to larger entropy production at reheating. As a result, more dilution of lepton asymmetry occurs for larger choice of $\xi$. This can be understood more clearly by noting that the scale factor corresponding to the onset of reheating epoch takes the form 
\begin{align}
A_{\rm RH}\simeq \left(\frac{45\,\alpha_k^\xi}{4\sqrt{3}\,\pi^2\,\gs(\Trh)}\right)^{1/4}\,\left(\frac{\rend}{\Trh^2\,M_P^2}\right)^{1/4}\,,    
\end{align}
following Eq.~\eqref{eq:rhoR} (for $k\gg 2$). Consequently, for larger $\Trh$ one would have smaller $A_{\rm RH}$, i.e., reheating ends earlier. In the bottom panel, we illustrate a scenario where the right baryon asymmetry is satisfied for $\xi=3$, with $\text{Re}[\theta]=0.76\,\pi,\,k=20$ and $\text{Im}[\theta]=1.76$.

Finally, it can be noted from the the top panel of Fig.~\ref{fig:scan} that the observed baryon asymmetry is achievable for any $k$ by properly tuning the parameters of the rotation matrix in CI prescription for the minimal gravitational reheating scenario. Here, we have scanned over real and imaginary values of the rotation parameter. We have considered the maximum value of $\text{Im}[\theta]=3.0$, in order to abide by the perturbative limit on the Yukwa coupling in Eq.~\eqref{eq:Ynu}. As seen from the figure, for a given RHN mass $M_1=10^{11}$ GeV, right baryon asymmetry is obtained for lower $\text{Im}[\theta]$ when the leptonic asymmetry is generated during the radiation-dominated background. Note that, due the condition $\Gamma_\tau^*>H$ at $T\sim M_1$, it is required to track the lepton asymmetry along the direction of $\tau$-flavour and along $(e+\mu)$-flavour, separately in this case. Consequently, here, 2FL framework is implemented to estimate the final $B-L$ asymmetry following~\cite{Nardi:2005hs,Nardi:2006fx,Datta:2022jic,Datta:2023pav}. On the contrary, for an inflaton-dominated background,  higher $\text{Im}[\theta]$ is required (along with the 1FL convention as discussed earlier) to obtain the correct baryon asymmetry. Additionally, with a larger $k$, smaller $\text{Im}[\theta]$ is found to be sufficient to produce correct asymmetry, that we have already noticed from Fig.~\ref{fig:yield-min1}. 

In the bottom panel of Fig.~\ref{fig:scan}, we explore the allowed parameter space for non-minimal gravitational reheating. To show the effect of non-minimal coupling $\xi$, here we fix $k=20$ and choose a few benchmark $\xi$-values that provide the observed baryon asymmetry. For comparison with the minimal case, we have shown the contour corresponding to $\xi=0$ via the purple dashed curve. As we increase $\xi$, we see, the allowed parameter space becomes identical to the one corresponding to radiation domination (shown by the green solid curve). This is expected since a larger $\xi$ results in a larger reheating temperature (see Fig.~\ref{fig:TRH}). Therefore, for larger $\xi$, more specifically, with $\xi>8$, reheating is also completed earlier (typically, above the scale of leptogenesis) and radiation domination begins. Consequently, on one hand, the $\tau$-Yukawa equilibration temperature $T_\tau^*$ becomes  identical to that of standard radiation-domination case, while on the other hand, the dominant contribution to leptonic asymmetry is sourced by the RHNs from thermal bath. As a result, the corresponding viable parameter space becomes similar to that of the RD scenario. 

We finally choose a RHN of mass $M_1=10^{10}$ GeV and show the corresponding allowed parameter space in Fig.~\ref{fig:scan2}. As already realized from the bottom panel of Fig.~\ref{fig:Tst}, with $M_1=10^{10}$ GeV and $k=16$, the $\tau$-lepton Yukawa equilibrates above the leptogenesis scale $T\sim 10^{10}$ GeV for $\xi\gtrsim 5$ leading to the 2FL prescription of leptogenesis. The resulting parameter space obtained by numerically solving the modified BEQ in Eq.~\eqref{eq:bl-flavor} is shown via the blue dashed contour in Fig.~\ref{fig:scan2}, where we consider the non-minimal coupling strength to be $\xi=6$ with $M_1= 10^{10}$ GeV and $k=16$. The 2FL regime for the standard radiation dominance scenario, obtained similarly by solving Eq.~\eqref{eq:bl-flavor}, is shown by the the green contour. To summarize, on increasing $\xi$, two things happen: (i) the lepton asymmetry shifts gradually from 1FL to 2FL regime, following Fig.~\ref{fig:Tst} and (ii) the parameter space eventually becomes identical to the scenario where the asymmetry is generated during radiation domination.
\section{Conclusion}
\label{sec:concl}
Gravitational interactions are irrefutable, and hence, gravity-mediated production of (beyond) the Standard Model (SM) particles is unavoidable. In this work, we therefore consider a scenario where the Universe is heated up at the end of reheating via gravitational production of the radiation bath. During the same time, we consider the production of heavy right-handed neutrinos (RHN), which further undergo CP-violating out-of-equilibrium decay to produce the observed baryon asymmetry via leptogenesis. These RHNs can be sourced either from the scattering of the inflaton condensate and SM particles in the thermal bath, mediated by graviton or inverse decay process from the thermal bath. However, the gravitational production from the thermal bath is always subdominant compared to the production from the inflation. Once the inflation gets over, the inflaton $\phi$ is assumed to oscillate in a monomial potential $\phi^k$. Depending on the choice of $k$, RHN mass, and the non-minimal coupling strength, it is possible that the charged lepton Yukawa interaction(s) equilibrates during reheating (typically $\tau$-flavour). We notice, for the minimal gravitational reheating case, the charged lepton Yukawa interactions remain out of equilibrium during the period of reheating. It is, therefore, safe to consider unflavoured leptogenesis in this situation. For non-minimal gravitational reheating, we found from Fig.~\ref{fig:Tst} that the $\tau$-lepton Yukawa may equilibrate, depending on the strength of the non-minimal coupling $\xi$ and the RHN mass, and in that case, the flavour effects must be taken into account (as depicted in Fig.~\ref{fig:scan2}). On the other hand, following Fig.~\ref{fig:scan} and ~\ref{fig:scan2}, we infer that as the non-minimal coupling becomes larger, the situation becomes identical to that of radiation domination as the reheating ends earlier, producing the radiation bath. 
\acknowledgments
We thank Debaprasad Maity and Arunansu Sil for useful discussions during the early stage of this project. The work of AD is supported by the National Research Foundation of Korea (NRF) grant funded by the Korean government (MSIT) (No. NRF-2022R1A4A5030362).
\appendix
\section{Details of Casas-Ibarra parametrization}
\label{sec:CI}
As the neutral component of the SM Higgs doublet acquires a VEV leading to the spontaneous breaking of the SM gauge symmetry, neutrinos in the SM obtain a Dirac mass that can be written as
\begin{align}
m_D= \frac{y_\nu}{\sqrt{2}}\,\langle H\rangle.
\end{align}
The Dirac mass $m_D$ together with the RHN bare mass $M_N$, can explain the nonzero light neutrino masses with the help of Type-I seesaw~\cite{GellMann:1980vs, Mohapatra:1979ia,MINKOWSKI1977421}. The light-neutrino masses can be expressed as,
\begin{align}
m_{\nu}\simeq -m_{D}\,M^{-1}\,m_{D}^T\,.
\end{align}
The mass eigenvalues and mixing are then obtained by diagonalising the light-neutrino mass matrix as
\begin{align}
D_m= \mathcal{U}^\dagger m_\nu \mathcal{U}^*
\end{align}
with $D_m={\rm dia}\,(m_1,m_2,m_3)$ consisting of the mass eigenvalues and $\mathcal{U}$ being the Pontecorvo-Maki-Nakagawa-Sakata matrix~\cite{Zyla:2020zbs}\footnote{The charged lepton mass matrix is considered to be diagonal.}. In order to obtain a complex structure of the Yukawa coupling which is essential from the perspective of leptogenesis, we use the well-known Casas-Ibarra (CI) parametrization~\cite{Casas:2001sr}. Using this one can write the Yukawa coupling $y_\nu$ as,
\begin{align}\label{eq:Ynu}
y_{\nu}=-i \frac{\sqrt{2}}{v} \mathcal{U} D_{\sqrt{m}} \mathcal{R}^T D_{\sqrt{M}},
\end{align}
where $\mathcal{R}$ is a complex orthogonal matrix $\mathcal{R}^T \mathcal{R} = I$, which we choose as
\begin{align}
\mathcal{R} =
\begin{pmatrix}
0 & \cos{z} & \sin{z}\\
0 & -\sin{z} & \cos{z}\\
1 & 0 & 0
\end{pmatrix}\,,
\label{eq:rot-mat}
\end{align} 
where $z=\text{Re}[\theta]+i\,\text{Im}[\theta]$ is a complex angle. The above structure of $\mathcal{R}$ can be justified by considering two RHNs or considering the third RHN $N_3$ to be very heavy and effectively decoupled from the bath. In such a scenario our neutrino Yukawa matrix becomes of dimension $3\times3$. Such a scenario also predicts the lightest active neutrino to be exactly massless. The diagonal light neutrino mass matrix $m_{\nu}^d$ is calculable using the best fit values of solar and atmospheric mass  obtained from the latest neutrino oscillation data~\cite{Zyla:2020zbs}. Now, the elements of Yukawa coupling matrix $y_N$ for a specific value of $z$, can be obtained for different choices of the heavy neutrino masses. For example, with $M_1=10^{11}$ GeV and $\{\text{Re}[\theta],\,\text{Im}[\theta]\}=\{0.76\pi,\,1.76\}$ (corresponding to the bottom panel of Fig.~\ref{fig:yield-min1}) we obtain the following structure
\begin{align}
y_\nu=\left(
\begin{array}{ccc}
 -0.0002 + 0.0076 i &  0.0696+0.00453i &  0 \\
 -0.0252-0.01072 i & -0.1014+0.2666 i & 0 \\
 -0.0142-0.0255 i & -0.241 + 0.151 i & 0
\end{array}
\right)\,,
\end{align}
which satisfies the light neutrino mass, as well as produces desired CP asymmetry in the visible sector, as we discuss below. 
\section{Non-minimal gravitational production}
\label{sec:non-minimal}
Here we derive the production rate of RHNs that stems from the non-minimal gravitational interaction. We begin with the action in the Jordan fame~\cite{Barman:2022qgt},
\begin{equation}
    \mathcal{S}_J = \int\,d^4 x \sqrt{-\tilde{g}} \left[-\frac{M_P^2}{2}\,\Omega^2\, \widetilde{\mathcal{R}} +\widetilde{\mathcal{L}}_{\phi}  + \widetilde{\mathcal{L}}_{h} + \widetilde{\mathcal{L}}_{N_i} \right]\,,
    \label{eq:jordan}
\end{equation}
where
\begin{align}
& \widetilde{\mathcal{L}}_{\phi} = \frac{1}{2}\,\partial_\mu  \phi\,\partial^\mu \phi-V(\phi)\\
& \widetilde{\mathcal{L}}_{h} =  \partial_\mu  h\,\partial^\mu h^\dagger -V(h h^\dagger) \\
& \widetilde{\mathcal{L}}_{N_i} = \frac{i}{2}\,\overline{\mathcal{N}_i}\,\overleftrightarrow{\slashed{\nabla}}\,\mathcal{N}_i  - \frac{1}{2}\,M_i\,\overline{(\mathcal{N})^c}_i\,\mathcal{N}_i + \widetilde{\mathcal{L}}_\text{yuk}\\
& \widetilde{\mathcal{L}}_\text{yuk} = -\left(y_\nu\right)_i\,
\overline{\mathbb{L}}\, h\,\mathcal{N}_i\,,
\end{align}
where $\mathcal{N}$ and $\mathbb{L}$ are the RHN and SM lepton doublet fields in Jordan frame. It is then possible to perform a conformal transformation from the Jordan frame to the Einstein frame,
\begin{equation}
    g_{\mu\nu} = \Omega^2\,\tilde{g}_{\mu\nu}\,,
\end{equation}
with $\Omega^2\equiv\left(1 + \xi_{\phi} \,\phi^2/M_P^2 + \xi_{h} \,|h|^2/M_P^2\right)$ being the conformal factor. The action in Einstein frame then becomes,
\begin{align}\label{eq:einstein0}
    & \mathcal{S}_E= \int d^4 x \sqrt{-g} \Biggl[-\frac{M_P^2\,\mathcal{R}}{2} + \frac{K^{ab}}{2}\,g^{\mu\nu} \partial_{\mu}S_a\,\partial_{\nu}S_b +  \frac{i}{2\,\Omega^3}\,\overline{N_i}\,\overleftrightarrow{\slashed{\nabla}}\,N_i-\frac{1}{\Omega^4}\,\left(\frac{M_i}{2}\,\overline{N_i^c}\,N_i+\mathcal{L}_\text{yuk}\right)  \nonumber\\&
    -\frac{3i}{4\Omega^4}\,\overline{N_i}\,\left(\overleftrightarrow{\slashed{\partial}}\,\Omega\right)\,N_i-\frac{1}{\Omega^4}\,\left(V_\phi+V_h\right) \Biggr]\,,
\end{align}
where
\begin{equation}
    \sqrt{-\tilde{g}} \rightarrow \frac{\sqrt{-g}}{\Omega^4} 
\end{equation}
\begin{equation}
    \tilde{\slashed{\nabla}} \rightarrow \Omega \slashed{\nabla} -\frac{3}{2}\Omega^2 (\slashed{\partial}\Omega)  \,, 
\end{equation}
and the indices $\{a,\,b\}\in\{\phi,\,h\}$. Following the field redefinitions: $L\rightarrow\Omega^{3/2}L$, $N_i \rightarrow \Omega^{3/2}N_i$ and $\overline{N_i}\rightarrow \Omega^{3/2}\overline{N_i}$,
we recover the action with canonical kinetic term for the RHNs as
\begin{align}
    \label{eq:einstein}
    & \mathcal{S}_E= \int d^4 x \sqrt{-g} \Biggl[-\frac{M_P^2\,\mathcal{R}}{2} + \frac{K^{ab}}{2}\,g^{\mu\nu} \partial_{\mu}S_a\,\partial_{\nu}S_b -  \frac{1}{\Omega^4}\,\left(V_\phi+V_h\right)+\frac{i}{2}\,\overline{N_i}\,\overleftrightarrow{\slashed{\nabla}}\,N_i 
    \nonumber\\&
    - \frac{1}{2\,\Omega}\,M_i\,\overline{N_i^c}\,N_i + \frac{1}{\Omega}\,\mathcal{L}_\text{yuk} \Biggr]\,,
\end{align}
where
\begin{equation}
    \label{eq:kinfunc}
    K^{ab}  = 6\,\frac{\partial \log \Omega}{\partial S_a} \frac{\partial \log \Omega}{\partial S_b} + \frac{\delta^{ab}}{\Omega^2}\,.
\end{equation}
The kinetic terms for the RHNs can be expressed in an effective form \begin{equation}
    \label{lag4point}
    \mathcal{L}_{\xi} \; = \; -\sigma_{hN_i}^{\xi}\, |h|^2\, \overline{N_i^c}N_i  - \sigma_{\phi N_i}^{\xi}\, \phi^2\,\overline{N_i^c}N_i \,,
\end{equation}
with 
\begin{equation}
    \label{appa:sigphiNi}
    \sigma_{\phi N_i}^{\xi}= \frac{M_i}{2M_P^2}\, \xi_{\phi}
\end{equation}
\begin{equation}
    \label{appa:sighNi}
    \sigma_{h N_i}^{\xi}=\frac{M_i}{2M_P^2}\,\xi_{h}\,.
\end{equation}
These non-minimal interactions provide additional production channels for the RHNs: $\phi\phi\to N_i\,N_i,\,hh\to N_i\,N_i$. Then, for the thermal production of RHNs, the production rate becomes,
\begin{align}\label{eq:rTxiN}
&  R^{T,\xi}_{N_i}\simeq N_h\,\frac{\zeta(3)^2\,\xi_h^2}{32\,\pi^5}\,\frac{M_i^2\,T^6}{M_P^4}\,, 
\end{align}
where $\zeta(x)$ is the Riemann-zeta function. For the inflaton scattering process $\phi\phi\rightarrow N_i\,N_i$, on the other hand, we find 
\begin{eqnarray}\label{eq:rphin}
    R^{\phi,\xi}_{N_i} &=& \frac{M_i^2\xi_{\phi}^2\phi_0^4\omega^2}{32 \pi M_P^4} \sum\limits_{n=1}^{\infty}(2n)^2|\mathcal{Q}^{(2)}_{2n}|^2\times  \sqrt{1-\frac{4M_i^2}{E_{2n}^2}}\,,
\end{eqnarray}
where we define $\phi_0=\left(\frac{\rho_\phi}{\lambda\, M_P^{4-k}}\right)^\frac{1}{k}$ and ${\cal Q}^{(2)}_n$ by
\begin{equation}
    \label{Eq:oscillation_2}
    \phi^2(t)=\phi_0^2(t)\cdot\mathcal{Q}^{2}(t)= \phi_0^2(t)\sum_{n=-\infty}^{\infty}\,{\cal Q}^{(2)}_n e^{-in \omega t}\,.
\end{equation} 
For small-field approximation ($\xi_\phi\ll 1$), $R^{\phi,\xi}_{N_i}$ is sub-dominant compared to Eq.~\eqref{eq:RphN}. Comparing with the minimal production rate in Eq.~\eqref{eq:RTN}, we find, the non-minimal thermal production dominates if
\begin{align}
\xi \gtrsim 5.36\,\frac{T}{M_1}\,,    
\end{align}
implying for $\xi\gtrsim 5.3$, the non-minimal contribution from thermal bath becomes important for $T\simeq M_1$. Similarly, comparing Eq.~\eqref{eq:RphN} and \eqref{eq:rTxiN}, we see, for
\begin{align}
& \xi \gtrsim  3.6\,\left(\frac{\rp}{m_\phi\,T^3}\right)\,,     
\end{align}
the non-minimal thermal contribution dominates over minimal non-thermal contribution. Now, since $\rho_\phi\gg m_\phi\,T^3$ for the parameter space of our interest, in this case, the non-minimal coupling needs to be $\gtrsim\mathcal{O}(10^{10})$ in order to have $ R^{T,\xi}_{N_i}\gtrsim R_{N}^\phi$.   
\bibliographystyle{JHEP}
\bibliography{Bibliography}
\end{document}